\documentclass[a4paper,11pt]{article}
\usepackage{jheppub} 
\usepackage[T1]{fontenc} 
\usepackage[normalem]{ulem}
\usepackage{threeparttable}
\usepackage{multirow}
\usepackage{multicol}
\usepackage{enumerate}
\usepackage{amssymb,bm,mathrsfs,bbm,amscd}
\usepackage{float}
\usepackage[section,above]{placeins}
\usepackage{units}
\usepackage{lineno}
\usepackage{subcaption}  
\def \ee   {e^{+}e^{-}}
\def \piz  {\pi^{0}}
\def \pip  {\pi^{+}}
\def \pim  {\pi^{-}}

\def \ifb  {\mbox{fb$^{-1}$}}
\def \gev  {\mbox{GeV}}
\def \gevc {\mbox{GeV/$c$}}
\def \gevcc{\mbox{GeV/$c^2$}}
\def \mev  {\mbox{MeV}}

\def \mevcc{\mbox{MeV/$c^2$}}

\def \mbc {M_{\rm{BC}}}
\def \dE {\Delta E}
\def \ebeam {E_{\mathrm{beam}}}

\def \Ks {K^0_{S}}

\def \modeTo be filled {\Sigma^{0}\pi^+}

\def \ModeTo be filled {\bar{\Sigma}^{0}\pi^-}

\def \lcp {\Lambda_{c}^{+}}
\def \lcm {\bar{\Lambda}_{c}^{-}}

\newcommand{\BESIIIorcid}[1]{\href{https://orcid.org/#1}{\hspace*{0.1em}\raisebox{-0.45ex}{\includegraphics[width=1em]{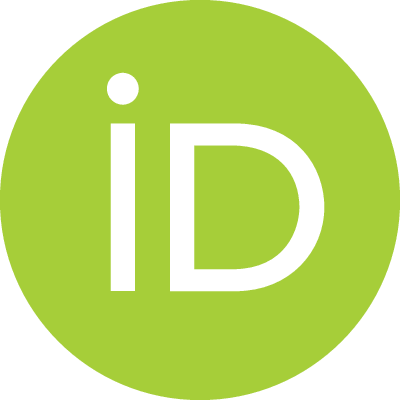}}}}

\arxivnumber{xxxx.xxxxx}

\title{\boldmath Amplitude analysis of the charmed baryon decay $\lcp \to p K_S^0\pi^0$}

\collaborationImg{\includegraphics[width=0.15\textwidth, angle=90]{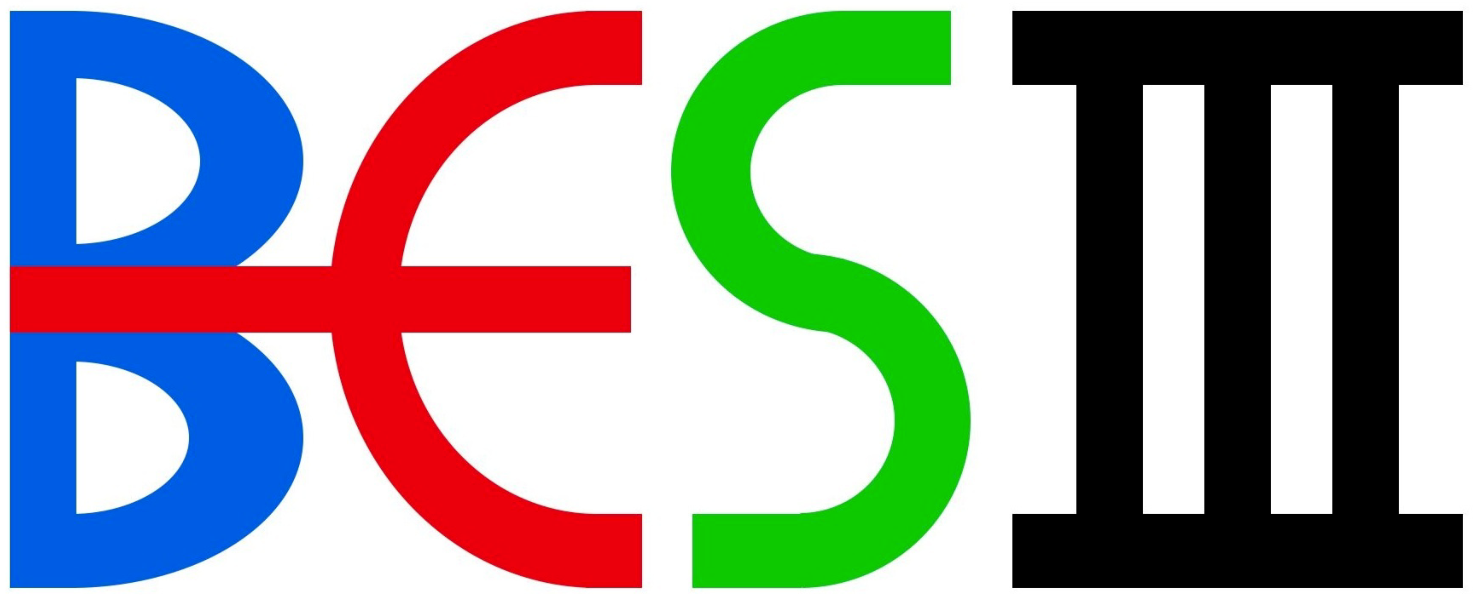}}
\collaboration{The BESIII collaboration}
\emailAdd{besiii-publications@ihep.ac.cn}

\abstract{ Using $e^+e^-$ annihilation data corresponding to an integrated luminosity of 4.5 fb$^{-1}$ collected with the BESIII detector at center-of-mass energies between 4.600 and 4.699 GeV, the first amplitude analysis of the charmed baryon decay $\Lambda_c^+ \to p K_S^0 \pi^0$ is performed. The processes $\Lambda_c^+ \to p \bar{K}^{*}(892)^0$, $\Lambda_c^+ \to \Delta(1600)^+ K_S^0$, and $\Lambda_c^+ \to N(1710)^+ K_S^0$ are identified with statistical significances of $12.8 \sigma$, $14.5 \sigma$, and $6.2 \sigma$, respectively. The mass and width of $\Delta(1600)^+$ are measured to be $(1520.4 \pm 8.8 \pm 37.7)~\text{MeV}$ and $(199.4\pm26.4\pm39.9)~\text{MeV}$, respectively. The mass and width of $N(1710)^+$ are measured to be $(1719.7 \pm 9.4  \pm 9.1)~\text{MeV}$ and $(72.0 \pm 18.0 \pm 15.0)~\text{MeV}$, respectively. The branching fractions of these processes are determined to be
\[
    \mathcal{B}\left[\Lambda_c^{+} \rightarrow \Delta(1600)^{+} \left(\rightarrow p \pi^0\right) K_S^0\right] = \left(8.10 \pm 0.80 \pm 5.57\right) \times 10^{-3},
\]
\[
    \mathcal{B}\left[\Lambda_c^{+} \rightarrow N(1710)^{+} \left(\rightarrow p \pi^0\right) K_S^0\right]  = \left(1.58 \pm 0.48 \pm 0.77\right) \times 10^{-3},
\]
\[
    \mathcal{B}\left[\Lambda_c^{+} \rightarrow p \bar{K}^{*}(892)^0 \left(\rightarrow K_S^0 \pi^0\right) \right] = \left(2.63 \pm 0.47 \pm 0.29\right) \times 10^{-3},
\]
where the first uncertainties are statistical and the second systematic.
}


\begin{document}
\maketitle
\flushbottom

\section{Introduction}
\label{sec:intro}
\hspace{1.5em}

As a key probe for studying strong and weak interactions in the charm sector, complementing the insights gained from charmed mesons, the $\Lambda_c^+$ baryon --- the lightest ground-state charmed baryon with spin-parity $J^P=1/2^+$ --- was first discovered by the Mark II experiment in 1979~\cite{Abrams1980}. Despite decades of subsequent research, a full understanding of its properties has remained challenging, primarily due to the complexity of three-quark systems~\cite{PhysRevD.46.1042} and limited experimental data. Over the past decade, the BESIII experiment has collected an integrated luminosity of \(4.5\,\mathrm{fb}^{-1}\) of \(e^+e^-\) annihilation data at center-of-mass (c.m.) energies between 4.600 and 4.699 GeV, enabling extensive experimental studies of \(\Lambda_c^+\) decays~\cite{BESIII:2022dxl,BESIII:2022ulv,Li:2025nzx,Hongjian2023}. Among these studies, amplitude analyses of multibody final states provide a powerful approach to investigating the underlying intermediate states by describing multibody decays in terms of sequential two-body decay processes. These include \(\Lambda_c^+ \to BX\), where \(B\) denotes a baryon state and \(X\) can be a pseudoscalar meson (e.g., \(\pi^0\)~\cite{XinhaiLambdapipi2022}), a vector meson (e.g., \(\rho(770)^+\)~\cite{XinhaiLambdapipi2022}), or a scalar meson (e.g., \(a_0(980)^+\)~\cite{PhysRevLett.134.021901}). Such analyses provide valuable insights into the decay dynamics of the \(\Lambda_c^+\) baryon, while the resulting experimental measurements also provide important inputs for data-driven theoretical investigations, such as the determination of strong-interaction phase shifts in weak nonleptonic decays~\cite{Wang:2024wrm}.

The three-body decay $\lcp\to p\Ks\piz$ can proceed via the intermediate processes $\Sigma^{*+}\to p\Ks$, $\Delta^{*+}\to p\piz$,
$N^{*+}\to p\piz$ and $\bar{K}^{*0}\to \Ks\piz$. Among the intermediate baryon
states, the \(\Delta(1600)^+\) resonance (\(J^P=3/2^+\)) has been extensively studied
through amplitude analyses of \(N\pi\) scattering
experiments~\cite{Hunt2019,Sokhoyan2015,Cutkosky1980,Hohler1979}. However, as
summarised in Tab.~\ref{tab:Delta1600}, these measurements exhibit significant
discrepancies, with the reported masses and widths differing by nearly 100
$\mevcc$. In addition to these experimental uncertainties, recent theoretical
studies based on Hamiltonian Effective Field Theory (HEFT) and lattice QCD
calculations suggest that the \(\Delta(1600)^+\) resonance may not be a simple three-quark excitation of
\(\Delta(1232)^+\), but rather a dynamically generated \(p\pi\) system arising
from strong rescattering effects~\cite{hockley2024understandingnaturedelta1600resonance}. Studies of nucleon excitations are crucial for
understanding non-perturbative QCD. Among them,  the \(N(1710)^+\) resonance (\(J^P=1/2^+\)) exhibits significant
discrepancies in its measured width~\cite{Hunt2019,Sokhoyan2015,Shklyar2013,Cutkosky1980,Hohler1979}, 
similar to those observed for the \(\Delta(1600)^+\) resonance, as summarised in
Tab.~\ref{tab:Delta1600}. The analysis of the decay $\Lambda_c^+ \to p
    \Ks\pi^0$ provides complementary information to $N\pi$ scattering data
and can help resolve the existing discrepancies in the resonance
parameters of the $\Delta(1600)^+$ and $N(1710)^+$.

\begin{table}[t]
    \begin{center}
        \small
        \begin{tabular}{c|c|c|c|c|c}
            \hline \hline
            \multicolumn{2}{c|}{\(\Delta(1600)^+\)} &
            \multicolumn{2}{c|}{\(N(1710)^+\)}      &
            \multirow{2}{*}{Reference}              &
            \multirow{2}{*}{Year}                                                                                                                  \\ 
            \cline{1-4}
            Mass (\mevcc)                           & Width (\mev)  & Mass (\mevcc)  & Width (\mev)  &                                      &      \\ 
            \hline
            $\approx 1570$                          & $\approx 250$ & $\approx 1710$ & $\approx 140$ & PDG~\cite{ParticleDataGroup:2022pth} &      \\
            1664 $\pm$ 16                           & 322 $\pm$ 46  & 1648 $\pm$ 16  & 195 $\pm$ 46  & Hunt~\cite{Hunt2019}                 & 2019 \\
            1520 $\pm$ 20                           & 235 $\pm$ 30  & 1715 $\pm$ 20  & 175 $\pm$ 15  & Sokhoyan~\cite{Sokhoyan2015}         & 2015 \\
            -                                       & -             & 1737 $\pm$ 17  & 368 $\pm$ 120 & Shklyar~\cite{Shklyar2013}           & 2013 \\
            1600 $\pm$ 50                           & 300 $\pm$ 100 & 1700 $\pm$ 50  & 90 $\pm$ 30   & Cutkosky\cite{Cutkosky1980}          & 1980 \\
            1522 $\pm$ 13                           & 220 $\pm$ 40  & 1723 $\pm$ 9   & 120 $\pm$ 15  & Hoehler\cite{Hohler1979}             & 1979 \\
            \hline \hline
        \end{tabular}
        \caption{Summary of the Breit-Wigner (BW) masses and widths of  \(\Delta(1600)^+\) and \(N(1710)^+\) reported by different experiments as compiled by the Particle Data Group (PDG)~\cite{ParticleDataGroup:2022pth}.}
        \label{tab:Delta1600}
    \end{center}
\end{table}

The $\bar{K}^{*}(892)^0$ state, with $J^P=1^-$, is one of the most
well-established vector mesons in the strange sector. Through amplitude
analysis, the contribution of $\bar{K}^{*}(892)^0$ can be identified and
quantified by extracting the two-body decay chain $\Lambda_c^+ \to
    p\bar{K}^{*}(892)^0$ followed by $\bar{K}^{*}(892)^0 \to \Ks\pi^0$. Previous
measurements of the branching fraction (BF) $\mathcal{B}(\Lambda_c^+ \to p
    \bar{K}^{*}(892)^0)$ have been performed primarily using the charged final state
 $\Lambda_c^+ \to p K^- \pi^+$~\cite{PhysRevD.108.012023}. Extending these measurements to the neutral final state $\Lambda_c^+ \to p K_S \pi^0$ provides complementary information. The neutral channel provides an independent experimental cross-check with systematic uncertainties largely uncorrelated with those of previous charged-channel measurements~\cite{PhysRevD.108.012023}, and the consistency of the $\bar{K}^{*0}$ contribution extracted from the two final states validates the analysis strategy and provides a test of the isospin structure of the decay. This can improve our understanding of the $\Lambda_c^+$ decay mechanisms.

In this paper, the first amplitude analysis of \(\Lambda_c^+ \to p \Ks \pi^0\) is
performed using  $\ee$ collision data at c.m. energies between 4.600
and 4.699 GeV~\cite{BESIII:2022dxl,BESIII:2022ulv}, corresponding to an
integrated luminosity of 4.5 $\ifb$. The masses and widths of the
\(\Delta(1600)^+\) and \(N(1710)^+\) resonances are measured and the product BFs of the
intermediate resonant processes are determined. Charge conjugate channels are
implied throughout this paper.

\section{BESIII experiment and Monte Carlo simulation}
\label{sec:detector}
\hspace{1.5em} The $\uchyph=0$BESIII detector~\cite{BESIII:2009fln}
records symmetric $\ee$ collisions provided by the $\uchyph=0$BEPCII
storage ring~\cite{Yu:2016cof}, which operates at c.m. energies
ranging from 1.85 to 4.95 GeV, with a peak luminosity of $1.1 \times
    10^{33}\;\mathrm{cm}^{-2}\mathrm{s}^{-1}$ achieved at $\sqrt{s} =
    3.773\;\mathrm{GeV}$. The $\uchyph=0$BESIII detector has collected
large data samples in this energy region~\cite{BESIII:2020nme}.  The
cylindrical core of the $\uchyph=0$BESIII detector covers 93\% of the
full solid angle and consists of a helium-based multilayer drift
chamber (MDC), a plastic scintillator time-of-flight system (TOF), and
a CsI(Tl) electromagnetic calorimeter (EMC), which are all enclosed in
a superconducting solenoidal magnet providing a 1.0 T magnetic
field~\cite{Huang:2022wuo}. The solenoid is supported by an octagonal
flux-return yoke with resistive plate counter muon identification
modules interleaved with steel.  The charged-particle momentum
resolution at $1\,\gevc$ is $0.5\%$, and the measurements of the energy deposited in the MDC (d$E/$d$x$)
resolution is $6\%$ for electrons from Bhabha scattering. The EMC
measures photon energies with a resolution of $2.5\%$ ($5\%$) at
$1\,\gev$ in the barrel (end cap) region.  The time resolution in the
TOF barrel region is 68 ps, while that in the end cap region was
initially 110 ps. The end cap TOF system was upgraded in 2015 using
multi-gap resistive plate chamber technology, providing a time
resolution of 60 ps~\cite{Li:2017jpg,Guo:2017sjt,Lange:2001uf}. Of the
data used in this analysis, 87\% was collected with the upgraded end cap TOF.

Simulated samples generated with \textsc{geant4}-based~\cite{GEANT4:2002zbu}
Monte Carlo (MC) software, including the geometric description of the BESIII
detector and the detector response~\cite{YouZhengYun2008,Liang:2009zzb,Huang:2022wuo}, are used to
determine detection efficiencies and estimate potential background
contributions. The simulation describes the spread of beam energy and the
initial state radiation (ISR) in $e^+e^-$ annihilations with the generator
\textsc{kkmc}~\cite{Jadach:2000ir,Jadach:1999vf}. The inclusive MC samples,
corresponding to about 40 times the number of events in the data samples,
include the production of $\lcp\lcm$ pairs, open charm processes, the ISR
production of vector charmonium(-like) states, and the continuum processes
incorporated in \textsc{kkmc}~\cite{Jadach:2000ir,Jadach:1999vf}. The known
decay modes are modeled with \textsc{evtgen}~\cite{Lange:2001uf,Ping:2008zz}
using BFs taken from the PDG~\cite{ParticleDataGroup:2022pth}, and the
remaining unknown charmonium decays are modeled with
\textsc{lundcharm}~\cite{Chen:2000tv,Lange:2001uf}. The final-state radiation
from charged final-state particles is incorporated using
\textsc{photos}~\cite{Richter-Was:1992hxq}. For the production of $e^+e^- \to
    \lcp\lcm$ events, the Born cross-section line shape is used from the BESIII
measurements~\cite{BESIII:2017kqg,BESIII:2023rwv}. At each c.m. energy, the
signal events for the process $\ee \to \lcp\lcm$ are generated according to a uniform phase space (PHSP) distribution, with the $\lcp$ decaying inclusively.

\section{Event Selection}
\label{sec:eventSelection}
\hspace{1.5em}
Utilizing $\Lambda_c^+\Lambda_c^-$ pairs produced near threshold in $e^+e^-$ collisions, a single-tag strategy~\cite{charmedhadronsBESIII} is adopted.
The $\Lambda_c^+ \to p K_S\pi^0$ candidates are then identified by selecting combinations of charged tracks and photons that fulfill the selection criteria outlined below.

Charged tracks are required to be within $|\cos\theta|<0.93$, where $\theta$ is
the polar angle defined with respect to the $z$-axis, which is the symmetry
axis of the MDC. The distance of closest approach to the interaction point (IP) along the $z$ axis ($V_z$) and in the perpendicular plane ($V_r$) must be less than 10 cm and 1 cm, respectively. Particle identification (PID) for charged tracks
combines d$E/$d$x$ and the
flight time in the TOF to form a likelihood value $\mathcal{L}(h)$ for each
hadron ($h$) hypothesis, where $h = p, K$, or $\pi$. Charged tracks are
identified as protons if the proton hypothesis has the highest likelihood
($\mathcal{L}(p) > \mathcal{L}(K)$ and $\mathcal{L}(p) > \mathcal{L}(\pi)$).

Photon candidates are reconstructed from showers that are not associated with
any charged track in the EMC. The energy deposited in the EMC is required to be
greater than 25 MeV in the barrel region ($|\cos\theta| < 0.80$), and greater
than 50 MeV in the end cap region ($0.86 < |\cos\theta| < 0.92$). The EMC time
difference from the event start time is required to be less than 700 ns, to
exclude electronic noise and showers unrelated to the event. The opening angle
between each shower and antiproton ($\bar{p}$) must be greater than $25^\circ$,
to suppress the background due to the annihilation of $\bar{p}$ with the detector
material. The $\piz$ candidates are reconstructed from pairs of photon candidates with
invariant mass $M(\gamma\gamma)$ in the range 0.115 $\gevcc$ $< M(\gamma\gamma) < $ 0.150 $\gevcc$. A kinematic fit is performed by constraining the $\gamma\gamma$ invariant mass ($M(\gamma\gamma)$) to the known $\piz$ mass. The four-momentum of the $\piz$ is then updated using the results of this 1C fit, and candidates with a good fit quality are retained for further analysis.

The $\Ks$ candidates are reconstructed from two oppositely charged tracks,
treated as $\pip$ and $\pim$, without imposing PID requirements. The tracks
must satisfy $V_z < 20$ cm, with no selection applied on $V_r$. The
invariant mass of the $\pip\pim$ pair $(M(\pip\pim))$ is required to be in the range
0.487 $\gevcc <M(\pip\pim) <$ 0.511 $\gevcc$. A loose condition is applied where the decay length $L$
from the IP of the candidate $\Ks$ must exceed one-tenth of the
resolution $\sigma_L$, $(L/\sigma_L > 0.1)$.

Based on studies of inclusive MC samples, the dominant background events for the signal mode \mbox{$\lcp \to p\Ks\piz$} are the processes \mbox{$\lcp \to \Lambda (\to p\pim) \pi^+ \piz$} and \mbox{$\lcp \to \Sigma^+ (\to p\piz) \pi^+ \pim$}. They are rejected by vetoing events with invariant mass $M(p\pim)$
($M(p\piz)$) in the intervals $1.11\,\gevcc < M(p\pim) <$
$1.12\,\gevcc$ ($1.17\,\gevcc < M(p\piz) < 1.20\,\gevcc$).
The $\lcp$ candidates are formed from all possible combinations of $p$, $\Ks$, and $\piz$ candidates satisfying the above selection criteria. If there are
multiple combinations of $\lcp$ in an event, we choose the candidate with the
smallest magnitude of the energy difference, defined as $\dE = E - \ebeam$,
where $\ebeam$ is the beam energy and $E$ is the reconstructed energy of the
$\lcp$ candidate in the \(e^+e^-\) rest frame. Furthermore, the requirement
$-0.1 < \dE < 0.1~\gev$ is imposed.

\begin{figure}[!tp]
    \centering
    \includegraphics[width=0.45\textwidth]{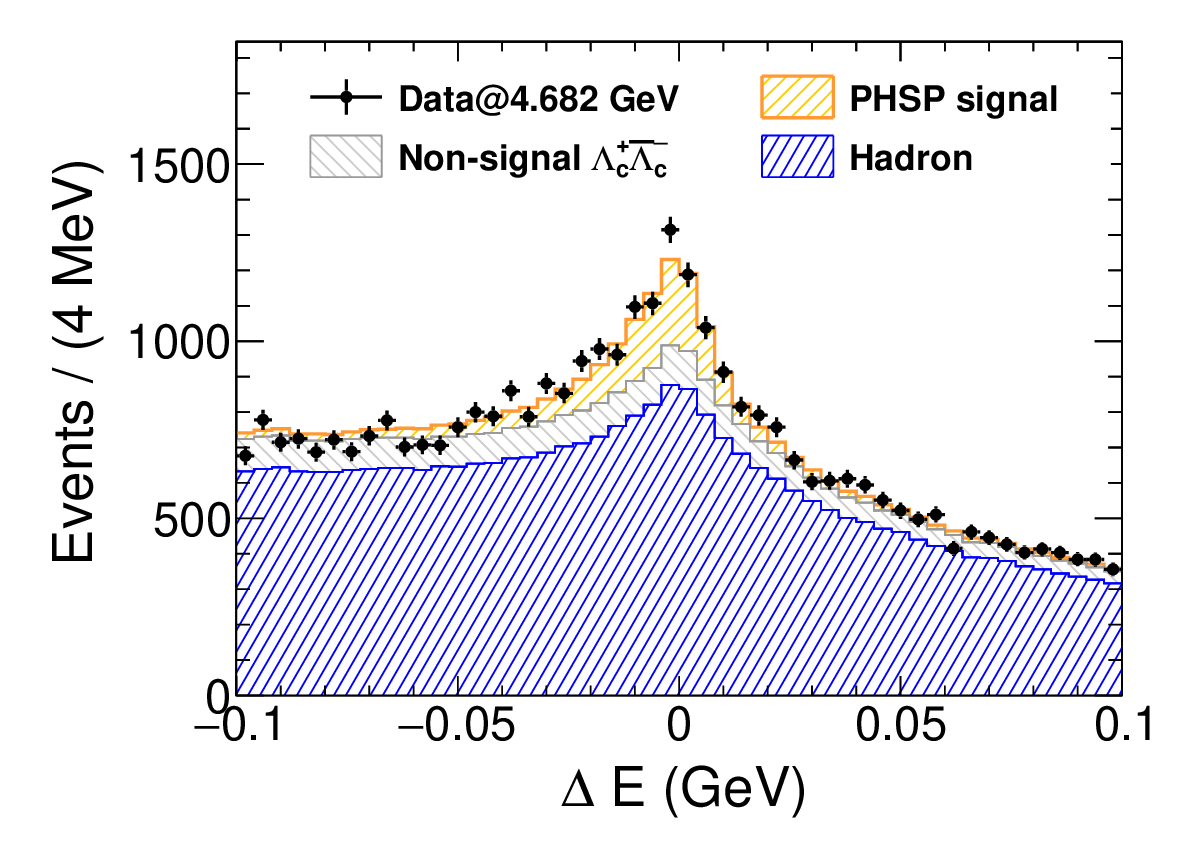}
    \includegraphics[width=0.45\textwidth]{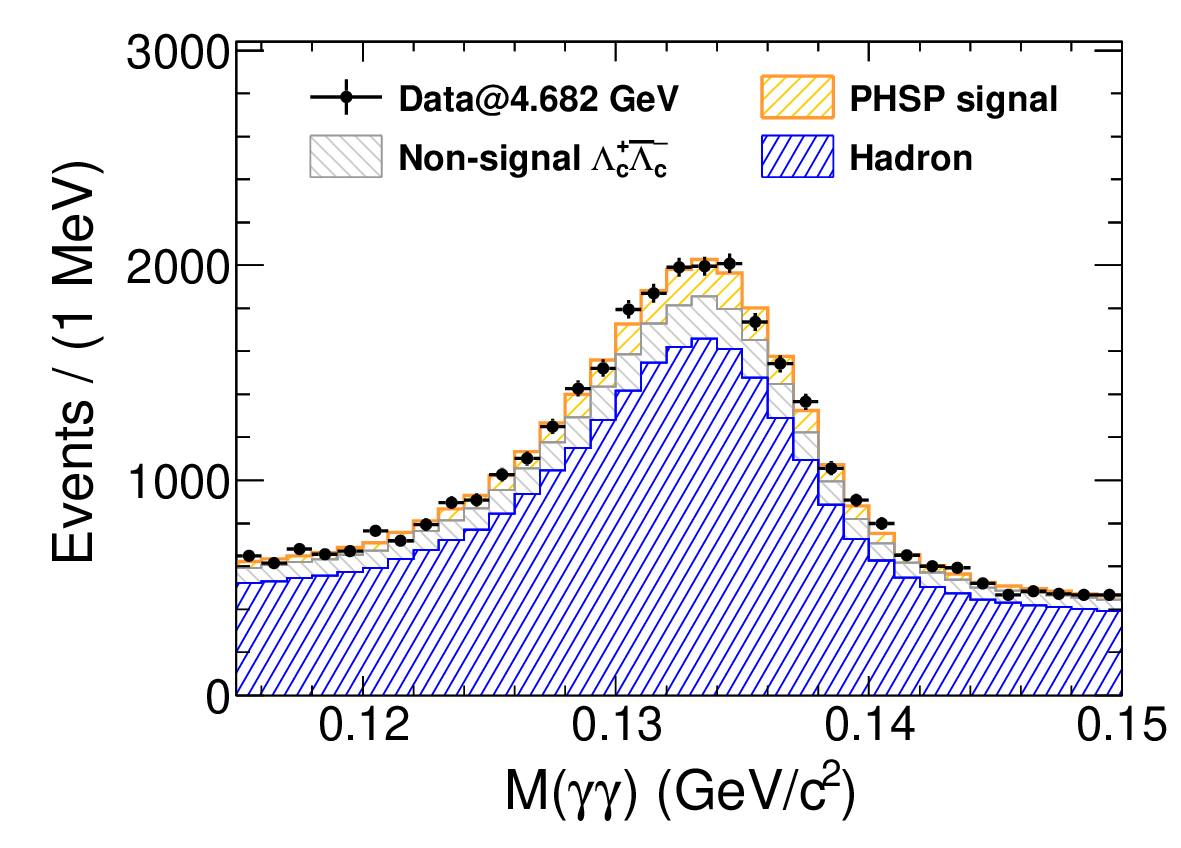}
    \includegraphics[width=0.45\textwidth]{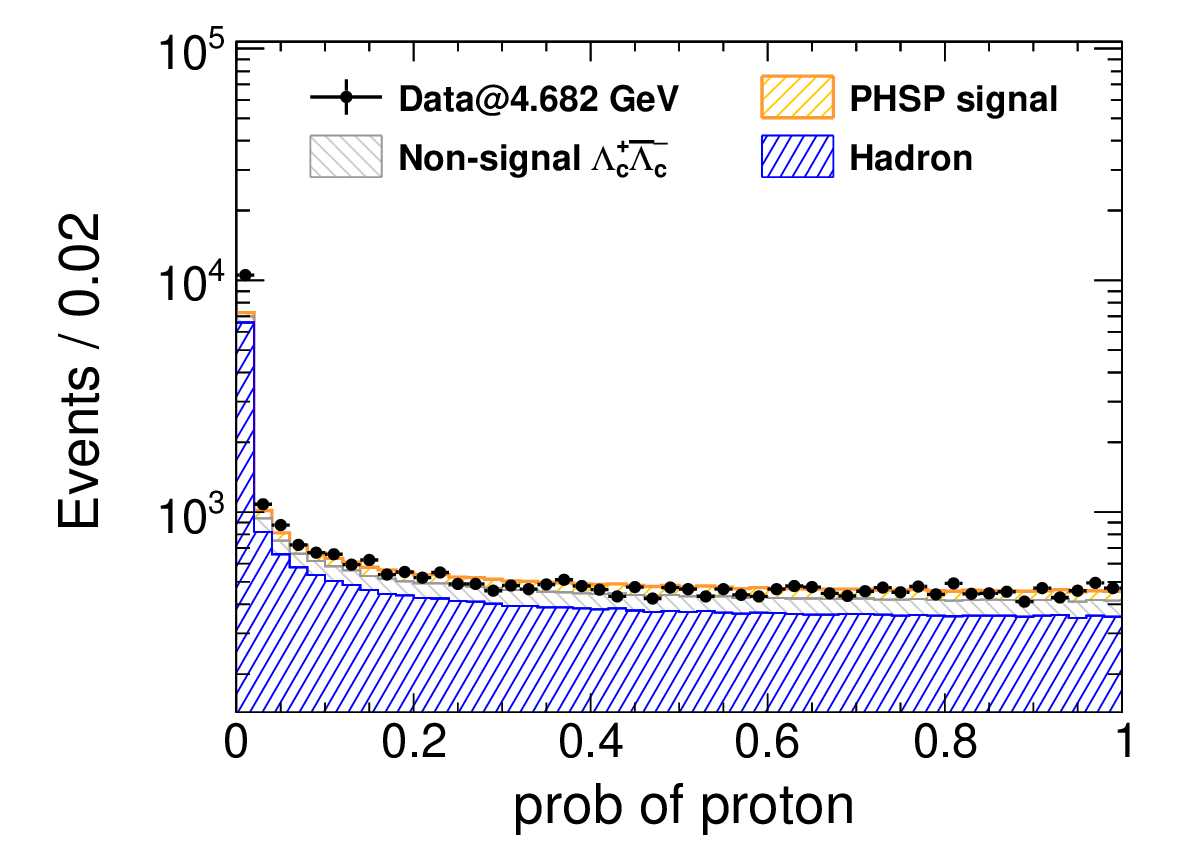}
    \includegraphics[width=0.45\textwidth]{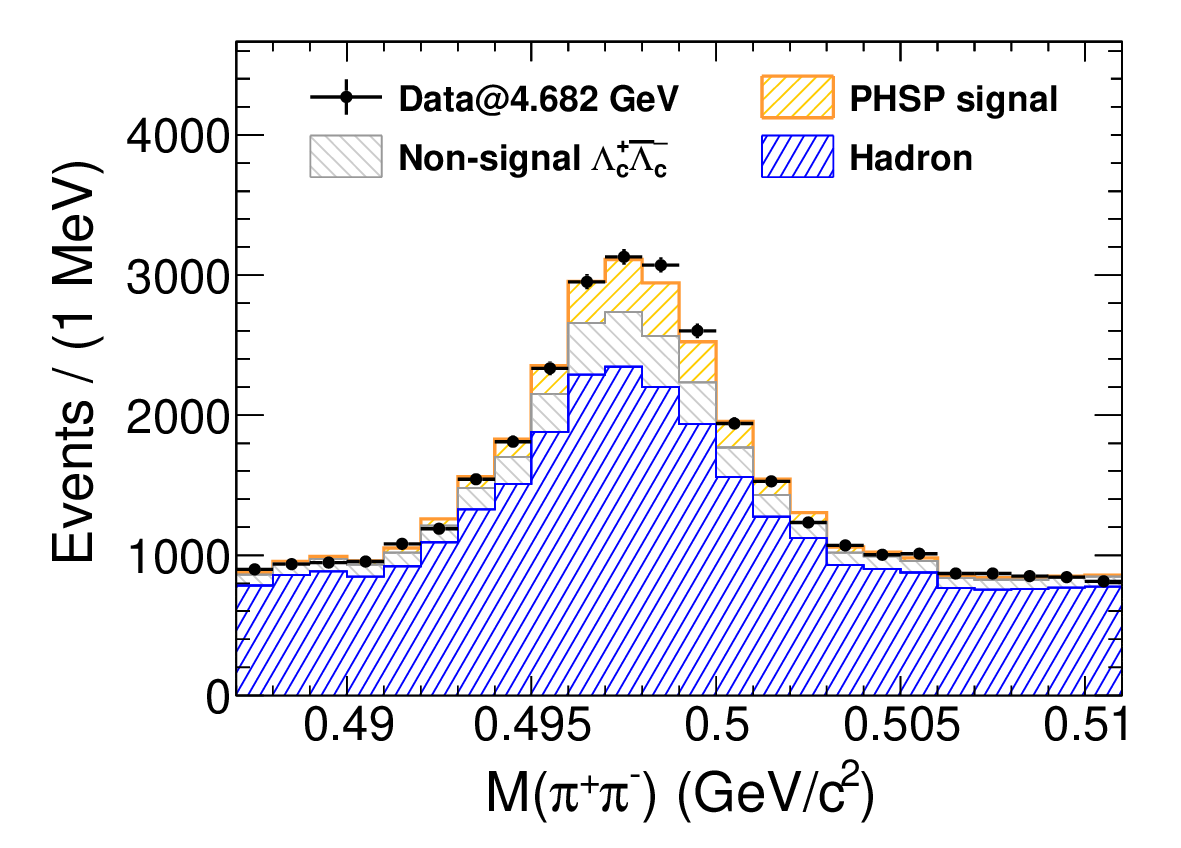}
    \includegraphics[width=0.45\textwidth]{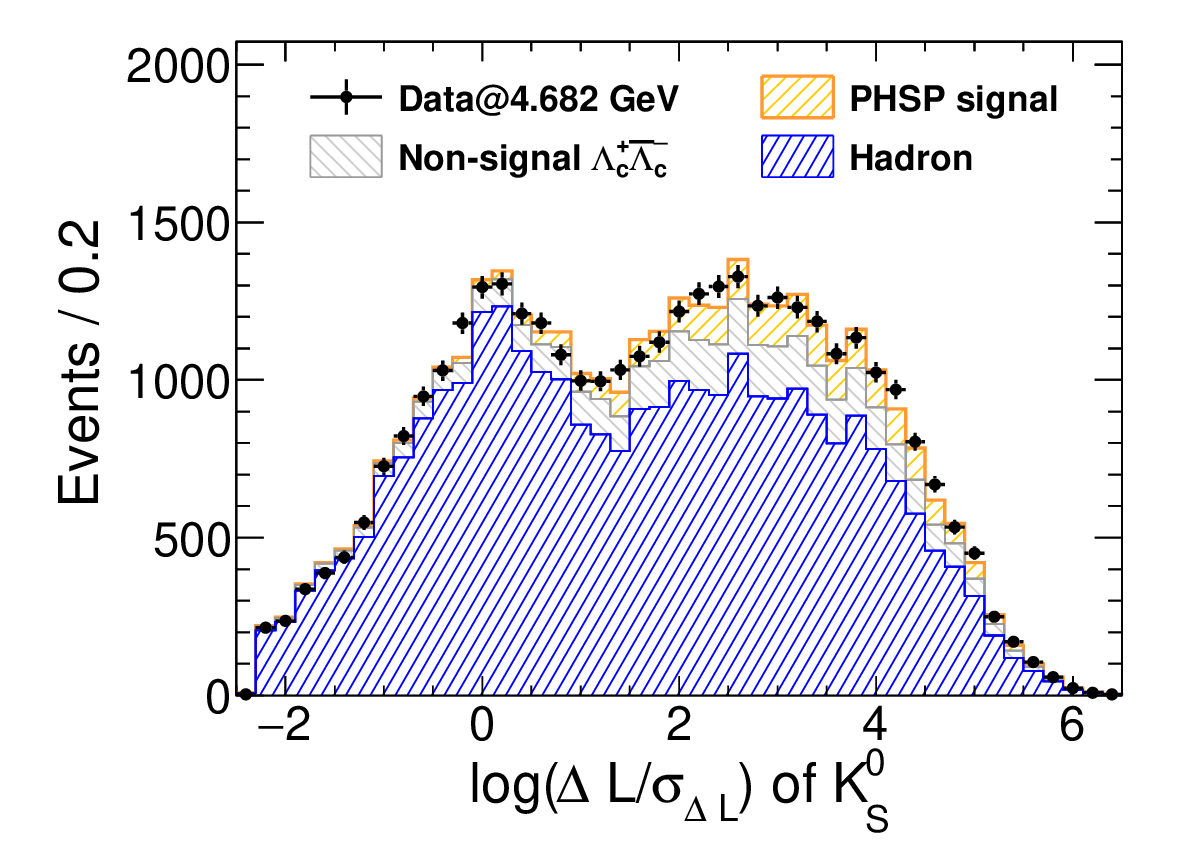}
    \includegraphics[width=0.45\textwidth]{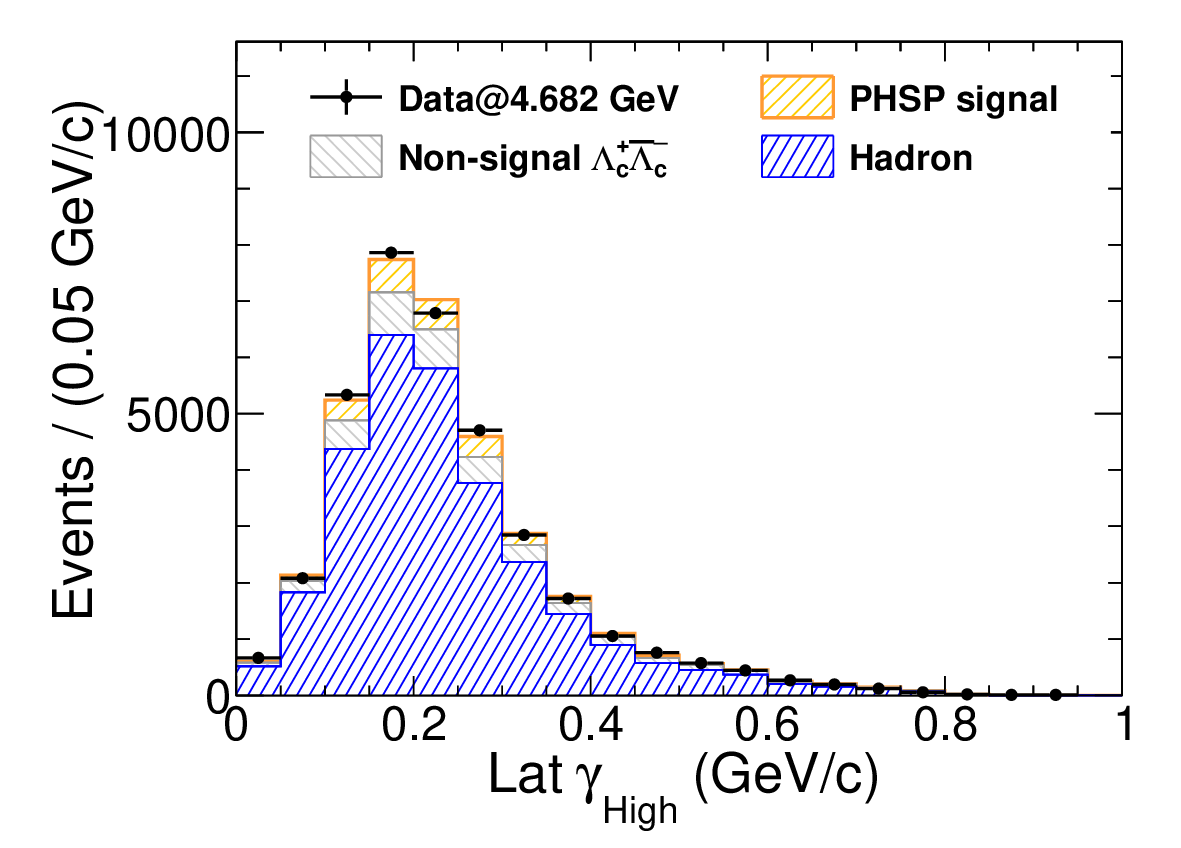}
    \includegraphics[width=0.45\textwidth]{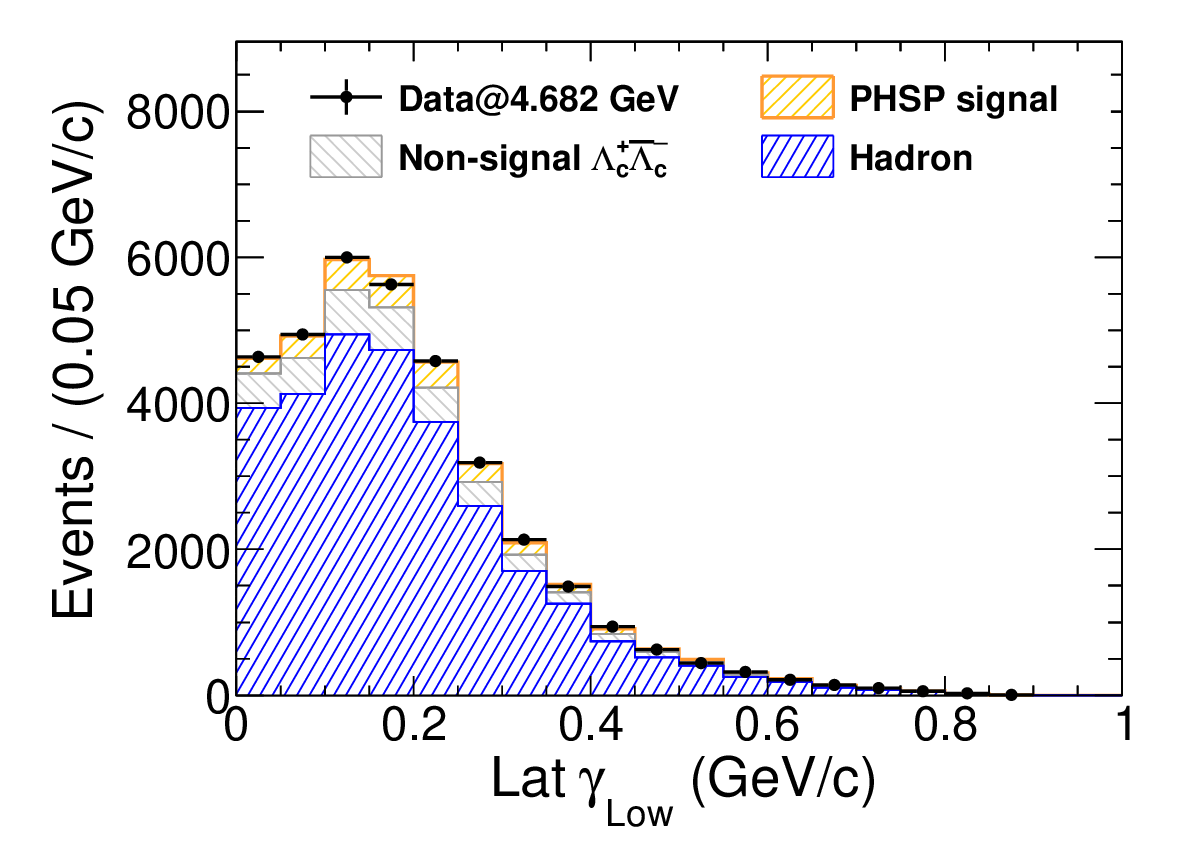}
    \includegraphics[width=0.45\textwidth]{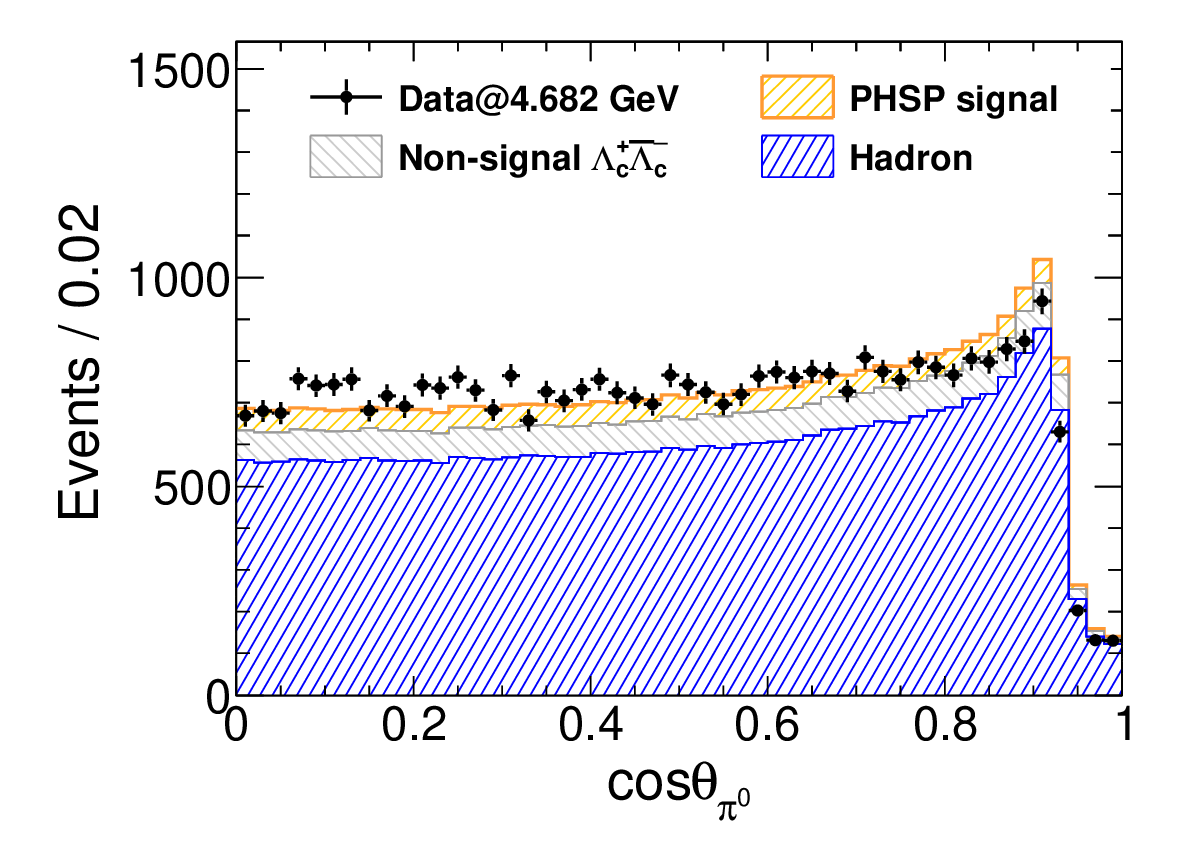}
    \caption{Distributions of training input variables for the BDTG at the c.m. energy of 4.682 \text{GeV}, with $M_{\text{BC}} > 2.20\,\text{GeV}$.}
    \label{fig:bdtg_inputs}
\end{figure}
The beam-constrained mass is defined as \(\mbc \equiv \sqrt{{\ebeam}^2/c^4 -
    \left|\overrightarrow{p}\right|^2/c^2}\), where \(\overrightarrow{p}\)
represents the total reconstructed momentum of the \(\lcp\) candidate in the
c.m. frame of the \(e^+e^-\) collision. To further suppress the background, a Gradient-Boosted Decision Tree (BDTG)
\cite{Hastie2009}, implemented using the TMVA package \cite{TMVA2009}, is used. The input variables are $\dE$ (whose distribution is already sculpted by the requirement of selecting the candidate with the smallest $|\dE|$ per event), $M(\gamma\gamma)$, $\mathcal{L}(p)$ of the proton, $M(\pip\pim)$, $\log \left(L / \sigma_L\right)$, the cosine of the helicity angle of $\piz\to\gamma\gamma$ ($\cos \theta_{\piz}$), the lateral moments of the showers corresponding to photons of higher energy $(\operatorname{Lat}\left(\gamma_{\text {High }}\right))$ and lower energy $(\operatorname{Lat}\left(\gamma_{\text {Low }}\right))$. In the BDTG training
sets, the signal inputs are obtained from the PHSP signal MC sample, and the
background inputs are obtained from non-signal processes in the inclusive MC
simulations. Background inputs are weighted to ensure consistency by comparing
the input variables between background MC simulations and data events in a
background-enriched region of the \( \mbc \) distribution, specifically \(
2.20\,\gevcc < \mbc < 2.25\,\gevcc \).

The c.m. energy of 4.682 GeV is chosen to illustrate the distributions of various variables due to its relatively high luminosity. As shown in Fig.~\ref{fig:bdtg_inputs}, the input variables of the training sets, after weighting the background events, show good agreement with the data at the c.m. energy of 4.682 GeV. 
The weights derived at $\sqrt{s}=4.682$ GeV are applied to all other c.m. energy points.
\FloatBarrier
\begin{figure}[!t]
    \centering
    \begin{subfigure}[t]{0.45\textwidth}
        \centering
        \includegraphics[width=\linewidth]{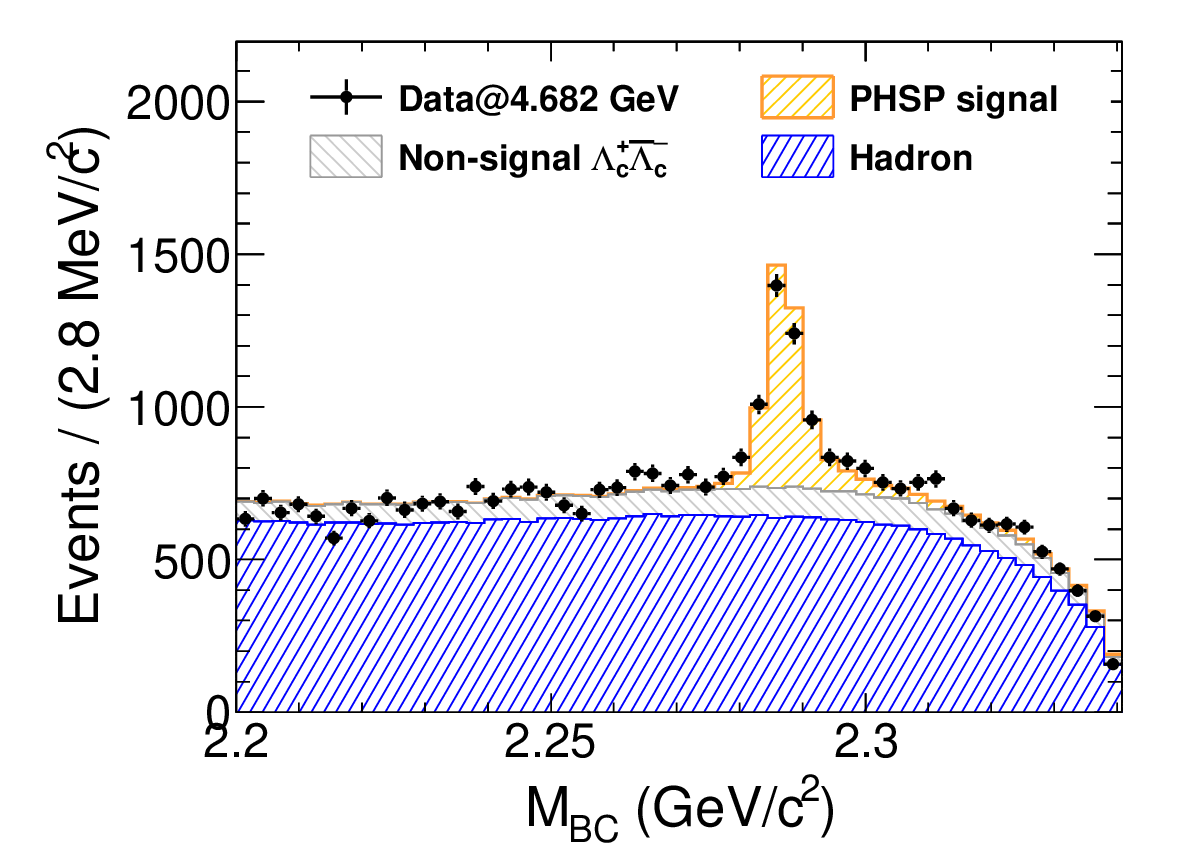}
        \caption{}
        \label{subfig:before_bdtg}
    \end{subfigure}
    \begin{subfigure}[t]{0.45\textwidth}
        \centering
        \includegraphics[width=\linewidth]{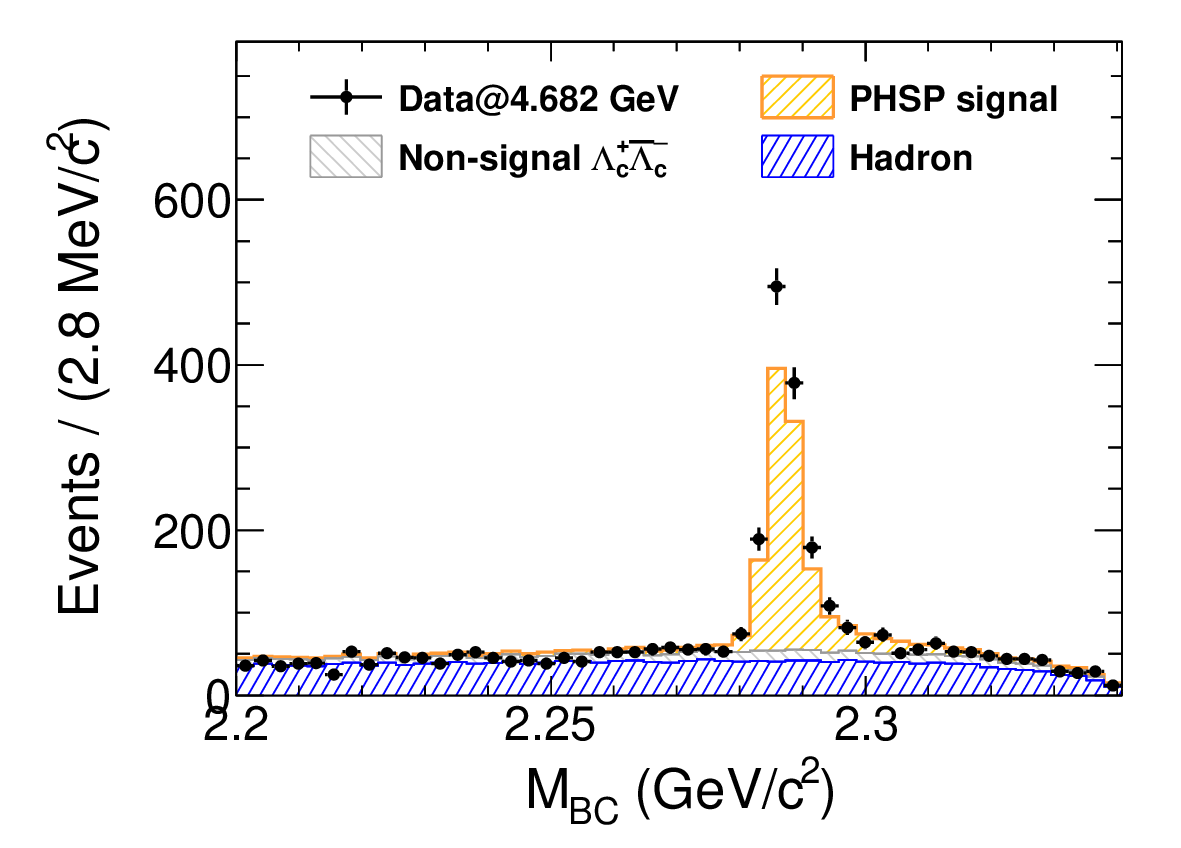}
        \caption{}
        \label{subfig:after_bdtg}
    \end{subfigure}
    \caption{The \(\mbc\) distributions at \(\sqrt{s} = 4.682~\text{GeV}\).
        Events passing (a) the first-stage event selection criteria and
        (b) further BDTG background suppression.}
    \label{fig:mbc_4682_comparison}
\end{figure}

 Fig.~\ref{subfig:before_bdtg} displays the \(\mbc\) distribution of the accepted candidates satisfying the aforementioned selection criteria, with background events weighted as described.

\FloatBarrier

\begin{figure}[!t]
    \centering
    \includegraphics[width=0.65\linewidth]{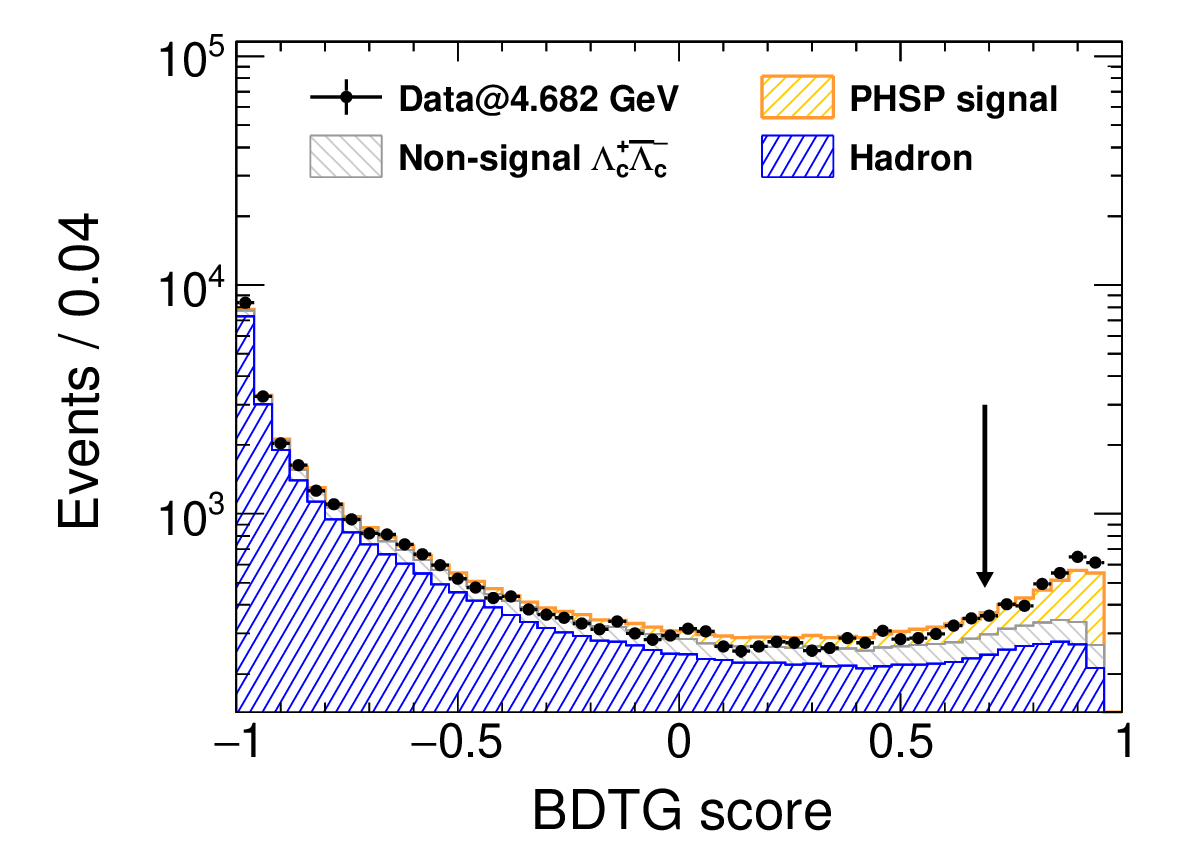}
    \caption{BDTG score distribution at the c.m. energy of 4.682\,\text{GeV}, with $M_{\text{BC}} > 2.20~\text{GeV}$. The arrow indicates the optimized BDTG cut value determined through the FOM scan. Events with a BDTG score above this threshold are selected as signal candidates.}
    \label{fig:BDTG_4682}
\end{figure}

Individual BDTG models are trained for each c.m. energy point. Different
requirements on the obtained BDTG score are then applied for the corresponding
energy point. To simultaneously optimize for both statistical significance and purity, the selection criteria for these BDTG scores are based on the figure-of-merit (FOM), defined as \mbox{\( \text{FOM} = \frac{S}{\sqrt{S+B}} \cdot \frac{S}{S+B} \)~\cite{PhysRevLett.131.031901, PhysRevLett.128.062001}}, where \( S \) (\( B \)) is the number of signal (background) events that pass the selection criteria. The FOM is calculated within the
signal region of \( \mbc \), where the left and right boundaries of the signal
region, as well as the cut on the BDTG score, are allowed to vary. The
optimization process converges after several iterations.
Fig.~\ref{fig:BDTG_4682} illustrates the distribution of the BDTG score at the c.m. energy of $4.682\,\text{GeV}$. The arrow indicates the optimized cut value of BDTG (listed in Tab.~\ref{tab:BDTG_MBc_Scan}), determined by the FOM scan. Events with a BDTG score above this threshold are selected as signal candidates. With this selection, a background rejection of $95.4\%$ in the signal region is achieved. The resulting $\mbc$ distribution at the c.m. energy of $4.682\,\text{GeV}$ is shown in Fig.~\ref{subfig:after_bdtg}.

\FloatBarrier

To determine the signal and background yields, an extended unbinned maximum
likelihood fit is performed on the $\mbc$ distribution, as shown in
Fig.~\ref{fig:fit_mbc}. In the fit process, the signal shape is derived
by convolving a smoothed PHSP MC histogram with a Gaussian function, to account for residual differences between simulation and data in resolution, beam-energy spread, and mass and energy scale.  The parameters of the Gaussian are left free in the fit. The background shape is modeled by an ARGUS function
\cite{ALBRECHT1993435}. The global probability density function comprises a
linear combination of signal and background contributions. The fit results in
the $\mbc$ signal regions are listed in Tab.~\ref{tab:BDTG_MBc_Scan}. The events in the $\mbc$ signal region, with a signal purity larger than $80\%$, are used for the subsequent amplitude analysis. To model the background contribution in the signal region, the $\mbc$ sideband data in the region $2.20\,\gevcc < \mbc < 2.27\,\gevcc$ are used to provide the background shape and normalization.

\begin{figure}[!tp]
    \centering
    \includegraphics[width=0.45\textwidth]{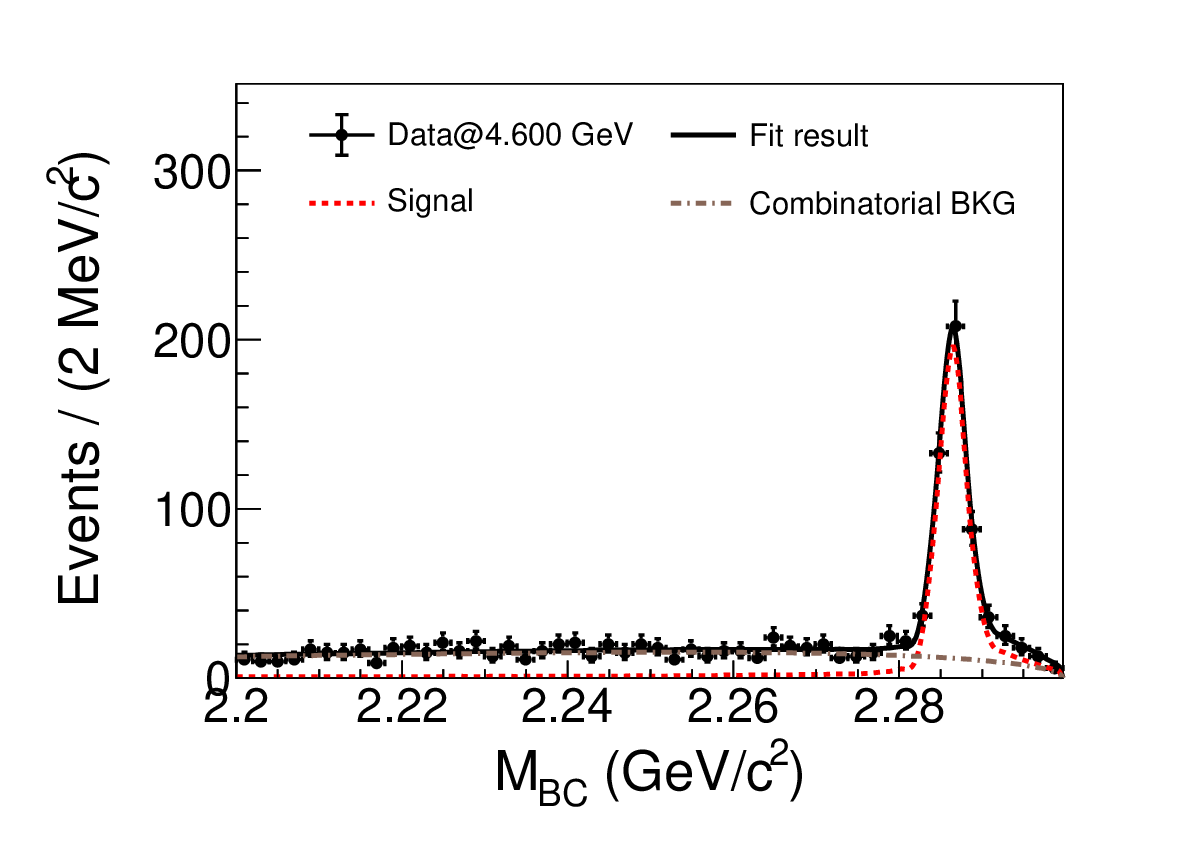}
    \includegraphics[width=0.45\textwidth]{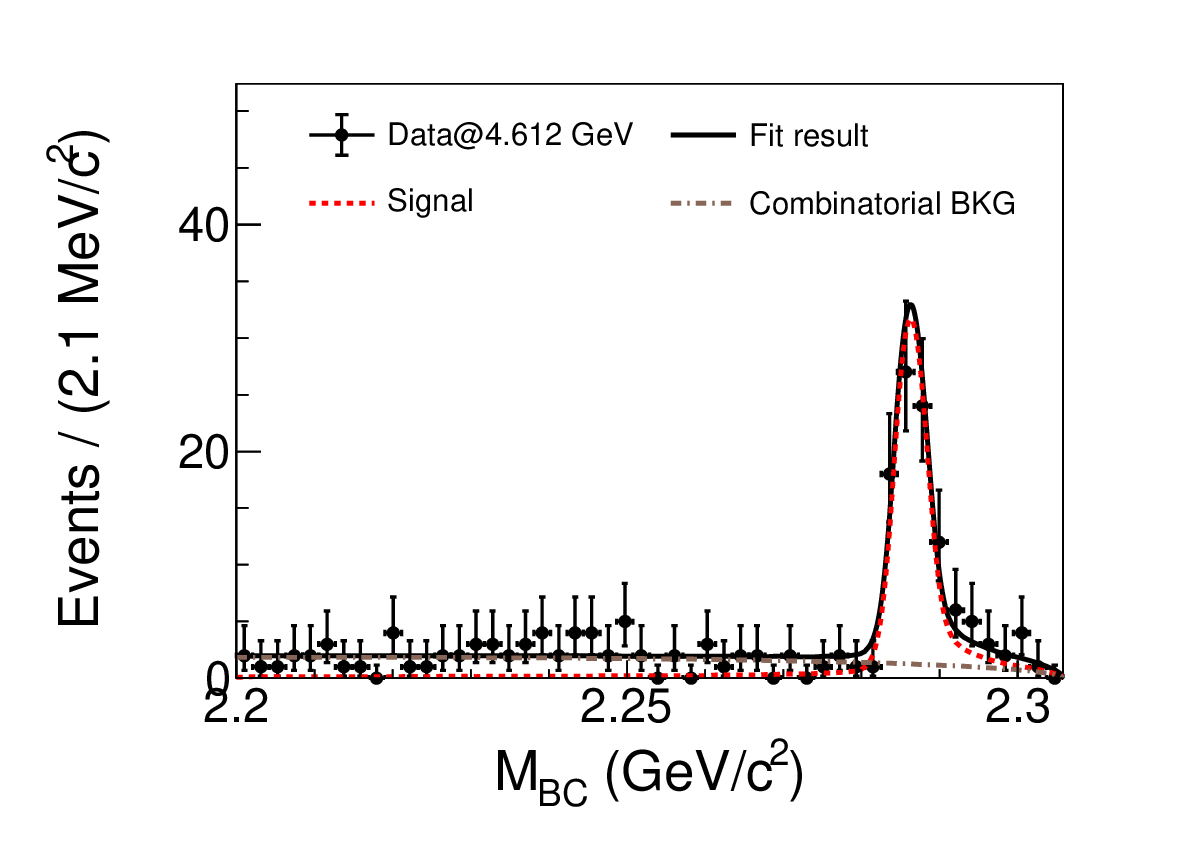}
    \includegraphics[width=0.45\textwidth]{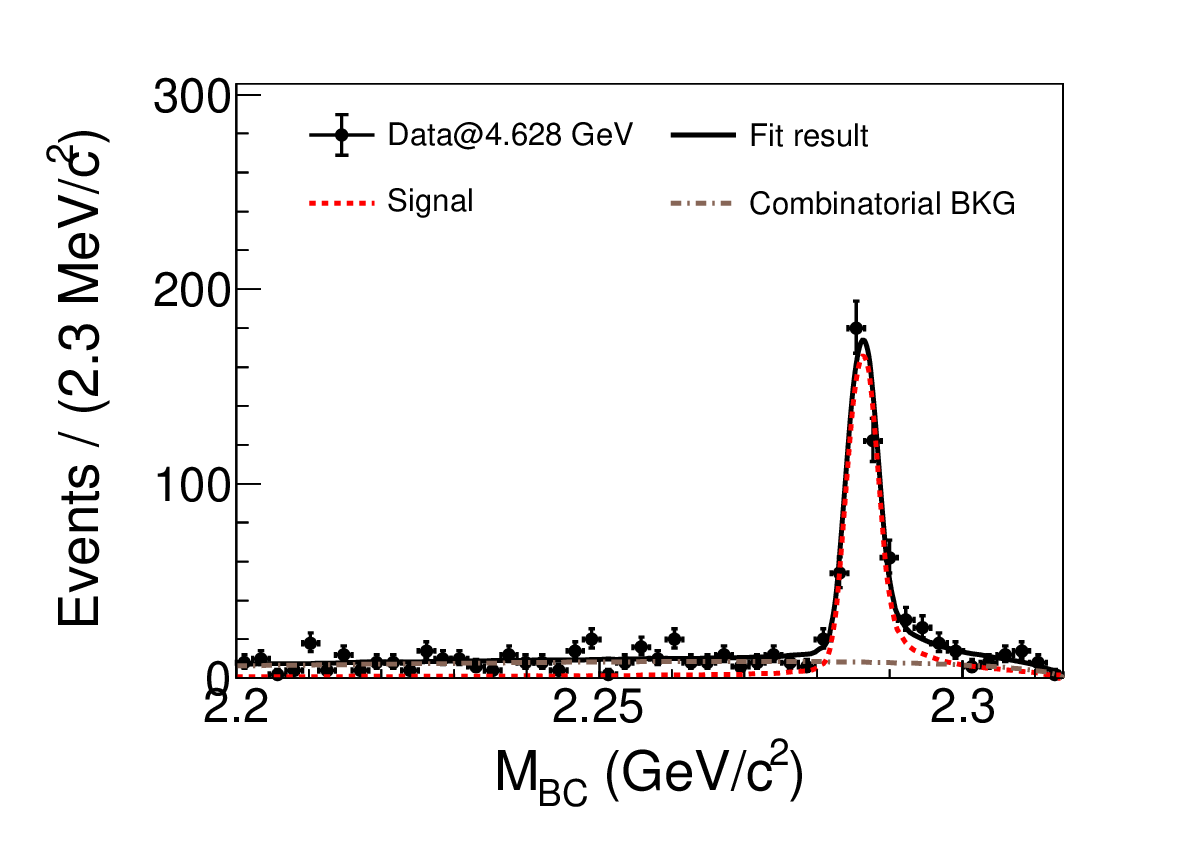}
    \includegraphics[width=0.45\textwidth]{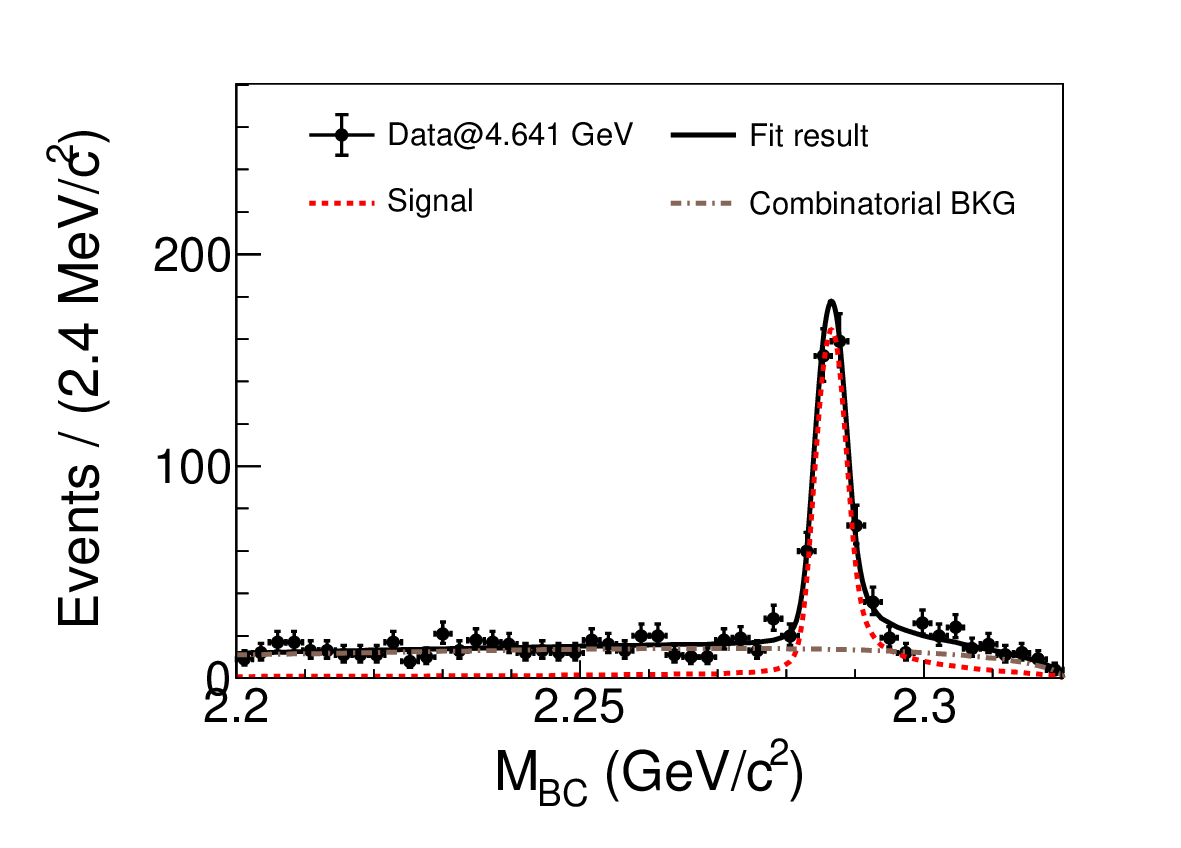}
    \includegraphics[width=0.45\textwidth]{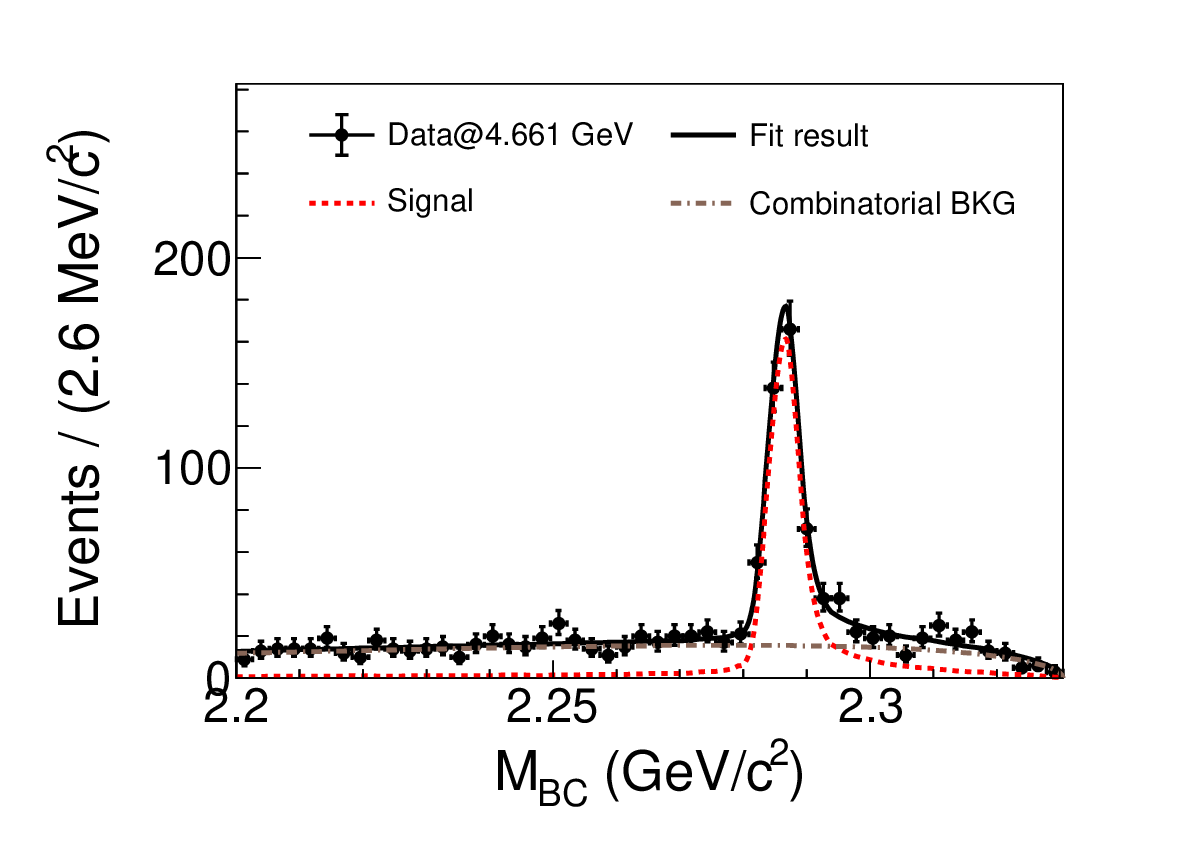}
    \includegraphics[width=0.45\textwidth]{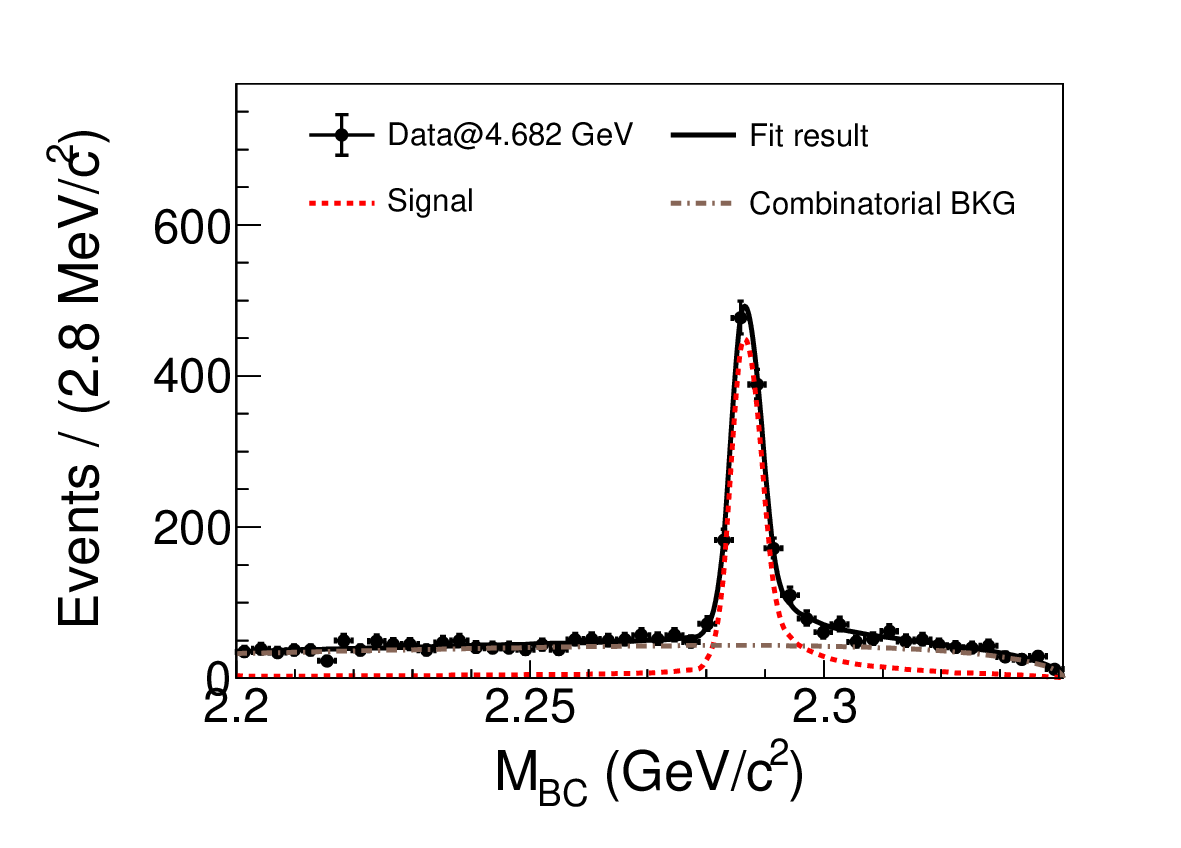}
    \includegraphics[width=0.45\textwidth]{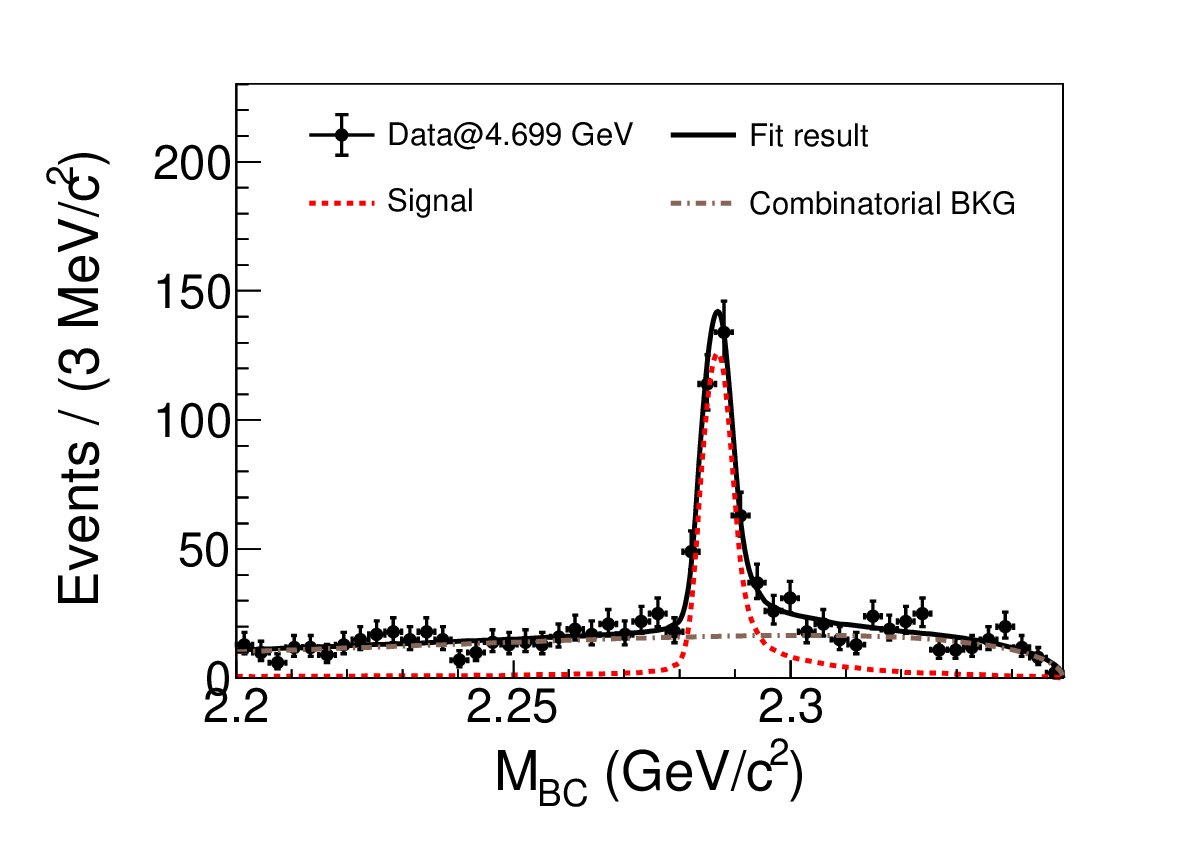}
    \caption{Fit results of the $\mbc$ distributions at seven c.m. energies.
        The black dots with error bars are data, and the black solid curves are the fit results. The red dashed curves are the fitted signal shapes. The brown dashed curves are the fitted combinatorial background shapes.}
    \label{fig:fit_mbc}
\end{figure}

\begin{table}[t]
    \centering
    \footnotesize
    \begin{tabular}{c|ccccc}
        \hline
        \hline
        $\sqrt{s}$($\mathrm{GeV}$) & \(M_{\text {BC }}\) signal region ($\gevcc$) & BDTG cut           & Signal yield  & Background yield & Purity $(\%)$   \\
        \hline
        4.600                      & $(2.283,2.291)$                             & $\textgreater0.50$ & $428 \pm 22$  & $46 \pm 2$       & $90.2 \pm 6.5$  \\
        4.612                      & $(2.283,2.292)$                             & $\textgreater0.64$ & $77 \pm 9$    & $5 \pm 1$        & $93.7 \pm 15.9$ \\
        4.628                      & $(2.283,2.291)$                             & $\textgreater0.50$ & $372 \pm 20$  & $29 \pm 2$       & $92.8 \pm 7.3$  \\
        4.641                      & $(2.283,2.292)$                             & $\textgreater0.50$ & $380 \pm 22$  & $51 \pm 3$       & $88.3 \pm 7.2$  \\
        4.661                      & $(2.283,2.291)$                             & $\textgreater0.50$ & $339 \pm 20$  & $48 \pm 2$       & $87.7 \pm 7.0$  \\
        4.682                      & $(2.282,2.293)$                             & $\textgreater0.69$ & $1048 \pm 36$ & $169 \pm 5$      & $86.1 \pm 4.1$  \\
        4.699                      & $(2.282,2.293)$                             & $\textgreater0.56$ & $291 \pm 19$  & $60 \pm 3$       & $82.9 \pm 7.4$  \\
        \hline
        Total                      & --                                          & --                 & $2935 \pm 58$ & $408 \pm 7$      & $87.8 \pm 0.3$  \\
        \hline\hline
    \end{tabular}
    \caption{Signal and background yields in the $\mbc$ signal region at each c.m. energy, along with $\mbc$ signal regions, BDTG cuts, signal purities. The uncertainties are statistical only.}
    \label{tab:BDTG_MBc_Scan}
\end{table}
\FloatBarrier

\begin{figure}[!t]
    \centering
    \includegraphics[width=0.45\textwidth]{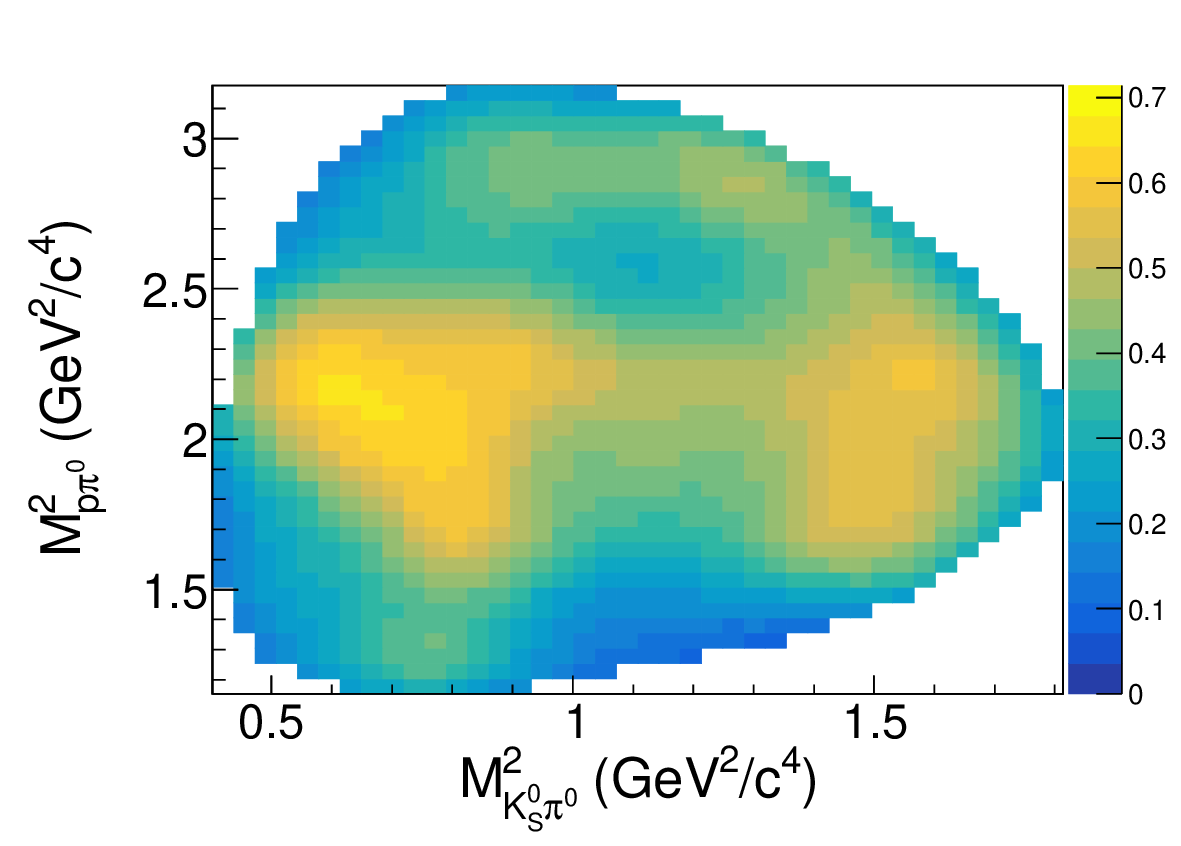}
    \includegraphics[width=0.45\textwidth]{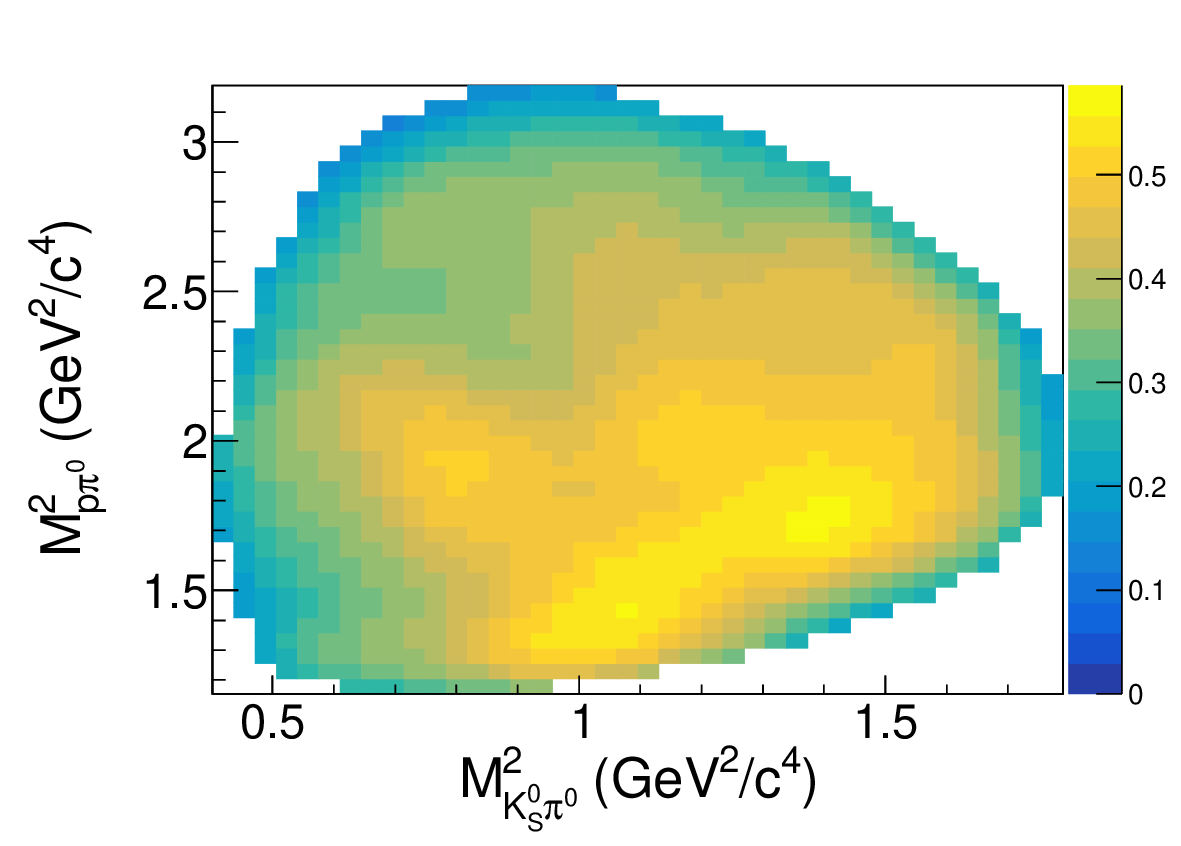}
    \caption{Dalitz plots for candidates in the $\mbc$ (left) signal and (right) sideband regions from the combined data at all c.m. energies.} 
    \label{fig:dalitzplot}
\end{figure}

\begin{figure}[!t]
    \centering
    \includegraphics[width=0.32\textwidth]{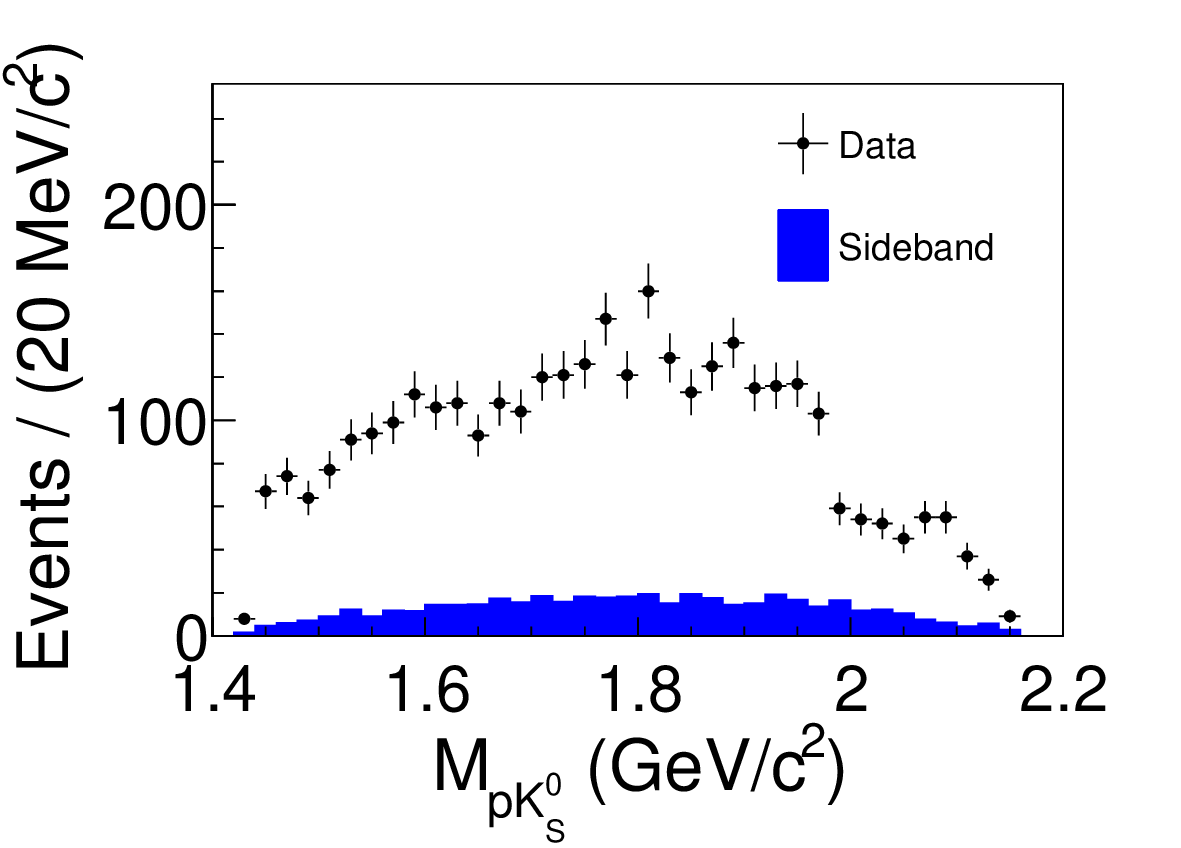}
    \includegraphics[width=0.32\textwidth]{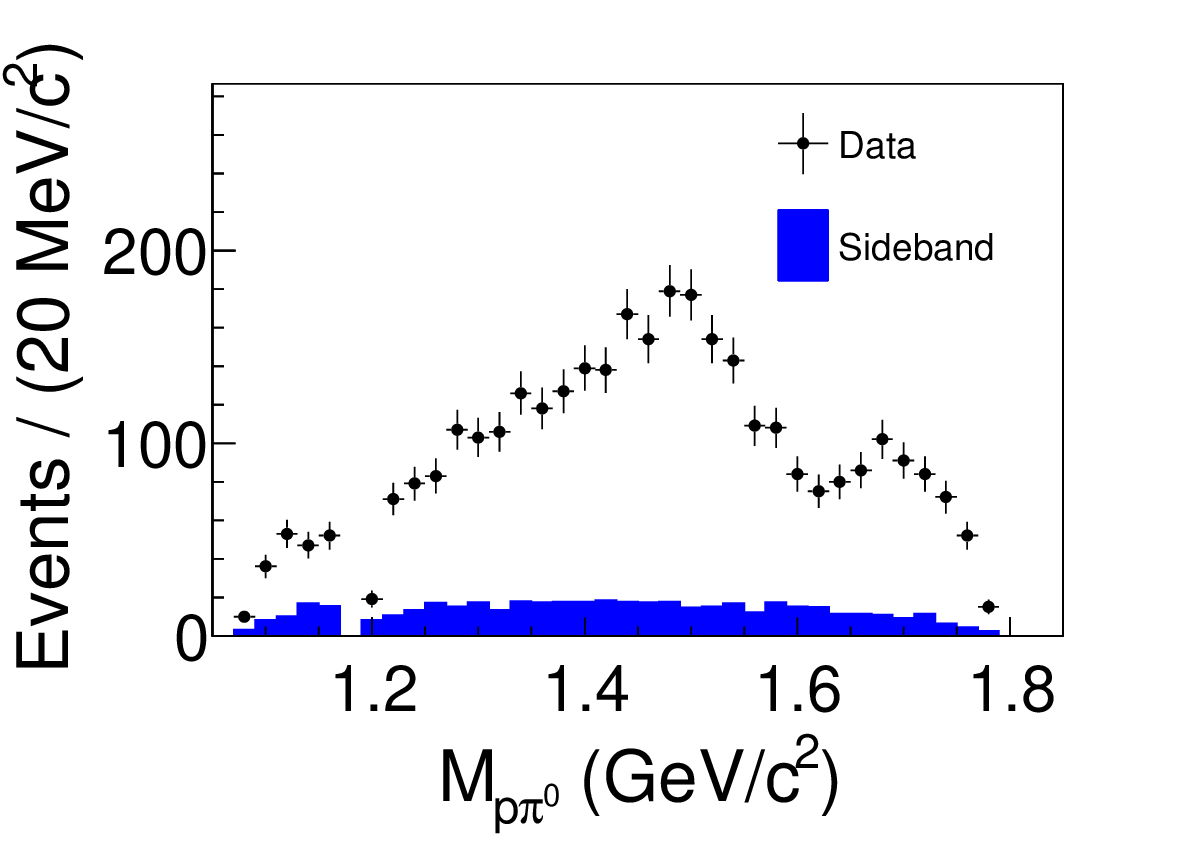}
    \includegraphics[width=0.32\textwidth]{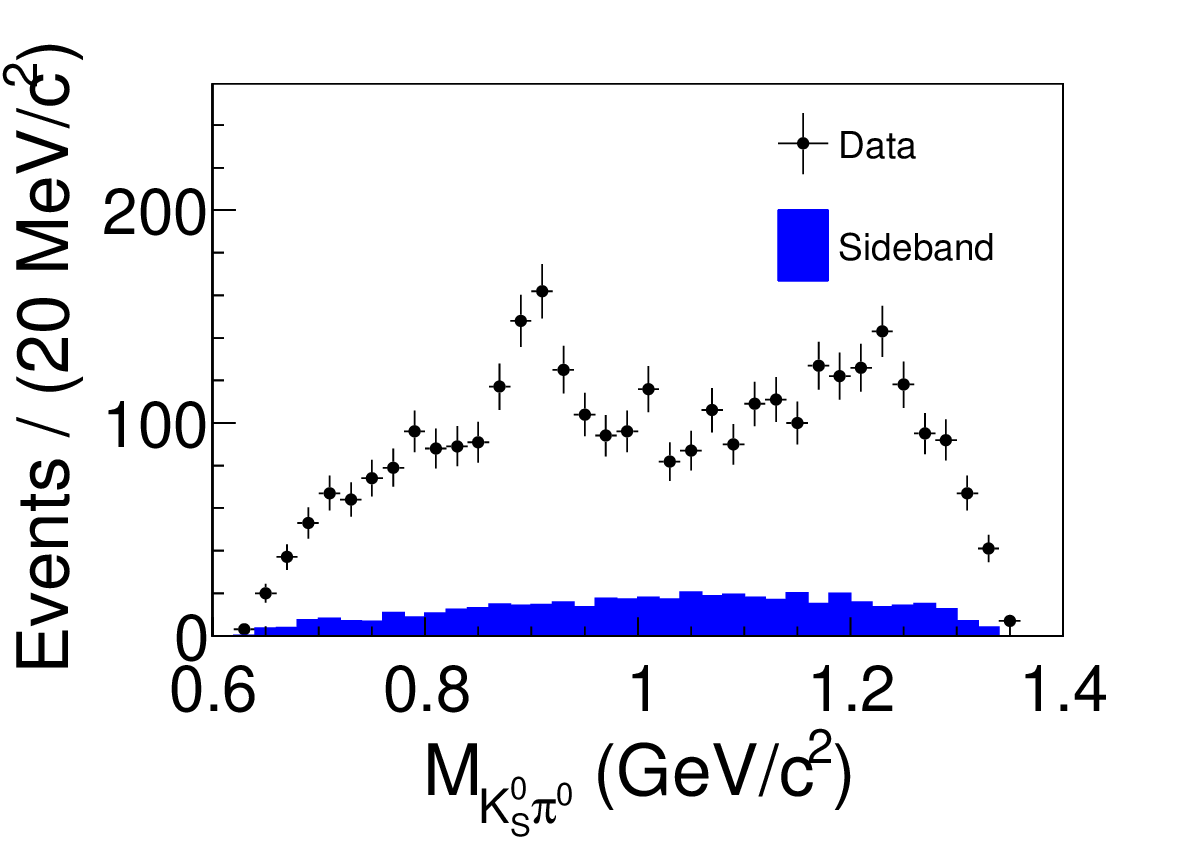}
    \caption{Distributions of (left) $M_{p\Ks}$, (middle) $M_{p\piz}$ and (right) $M_{\Ks\piz}$. The black points with error bars represent the data from the $\mbc$ signal region, while the blue histograms show the background shape extracted from the $\mbc$ sideband region. The background distributions are normalized 
to the fit results listed in Tab.~\ref{tab:BDTG_MBc_Scan}.}
    \label{fig:massplot}
\end{figure}

To improve momentum resolution, an additional kinematic fit is performed, in which the $p \Ks\piz$ system and the recoiling $\lcm$ are
constrained to the known $\lcp$ mass~\cite{ParticleDataGroup:2022pth}, the
$\pip\pim$ system is constrained to the known $\Ks$
mass~\cite{ParticleDataGroup:2022pth} and the $\gamma\gamma$ system is
constrained to the known $\piz$ mass~\cite{ParticleDataGroup:2022pth}. The $\lcm$ recoil momentum is calculated using the energy of the initial $\ee$ system and the momentum of
$p\Ks\piz$. The updated kinematic information of the signal candidates from the
kinematic fit is used in the amplitude analysis. The Dalitz plots for the combined
data for all c.m. energies, using the updated momenta, in the $\mbc$ signal and sideband regions are shown in Fig.~\ref{fig:dalitzplot}. The
two-body invariant mass distributions from the $\mbc$ signal region are shown in
Fig.~\ref{fig:massplot}.

\clearpage
\section{Amplitude analysis}
\label{sec:amp_ana}

\subsection{Benchmark fit procedure}
\hspace{1.5em} The amplitude analysis is performed using the open-source framework TF-PWA \cite{Jiang:2024vbw}, based on the helicity amplitude formalism described in Refs.~\cite{Chung1971,Richman1984}. The fundamental concepts follow Ref. \cite{Xinhai2022}. The three-body decay $\lcp\to p\Ks\piz$ is described as a sequential quasi-two-body decay within the isobar model, factorized into four subsystems: $\Sigma^{*+}\to p\Ks$, $\Delta^{*+}\to p\piz$, $N^{*+}\to p\piz$ and $\bar{K}^{*0}\to \Ks\piz$. For each two-body decay $0\to 1+2$, the helicity amplitude is written as
\begin{equation}
    A_{\lambda_0, \lambda_1, \lambda_2}^{0 \rightarrow 1+2}=H_{\lambda_1, \lambda_2}^{0 \rightarrow 1+2} D_{\lambda_0, \lambda_1-\lambda_2}^{J_0 *}(\phi, \theta, 0),
\end{equation}
where $D_{\lambda_0, \lambda_1-\lambda_2}^{J_0 *}$ is the Wigner $D$-function, $\phi$ and $\theta$ are the helicity angles. The amplitude $H_{\lambda_1, \lambda_2}^{0 \rightarrow 1+2}$ is given by the LS coupling formula \cite{Chung1997, Chung1998,Chung2003} along with the barrier factor terms
\begin{equation}
    H_{\lambda_1, \lambda_2}^{0 \rightarrow 1+2}=\sum_{l s} g_{l s} \sqrt{\frac{2 l+1}{2 J_0+1}}\left\langle l 0, s \delta \mid J_0, \delta\right\rangle\left\langle J_1 J_2, \lambda_1-\lambda_2 \mid s, \delta\right\rangle\left(\frac{q}{q_0}\right)^l B_l^{\prime}\left(q, q_0, d\right),
    \label{gls_def}
\end{equation}
where $g_{l s}$ is the complex-valued partial wave amplitude, and the two-body-decay helicity amplitude $A_{\lambda_0, \lambda_1, \lambda_2}^{0 \rightarrow 1+2}$ is constructed from $g_{l s}$; $J_{0,1,2}$ are the spins of the particles 0, 1 and 2, $\lambda_{1,2}$ are the helicities for the particles 1 and 2, and $\delta=\lambda_1-\lambda_2$ is the helicity difference. Here, $q$ is the three-momentum modulus of particle 1 in the rest frame of particle 0, which is calculated as
\begin{equation}
    q=\frac{\sqrt{\left[m^2-\left(m_1+m_2\right)^2\right]\left[m^2-\left(m_1-m_2\right)^2\right]}}{2 m},
\end{equation}
where $m, m_1$ and $m_2$ are the masses of the particles 0, 1 and 2, respectively. The normalization factor $q_0$ is calculated at the nominal resonance mass. The factor $B_l^{\prime}\left(q, q_0, d\right)$ is the reduced Blatt-Weisskopf barrier factor described in Refs. \cite{BlattWeisskopf1952, HippelQuigg1972}, and the radius of the centrifugal barrier $d$ is chosen as $d = 3.69~(\gevc)^{-1}$ ($0.73~\text{fm}$) which is the same as in Ref. \cite{Xinhai2022}.

The amplitude of a complete decay chain is constructed as the product of each
two-body decay amplitude and the resonant propagator $R$. For the decays
through the $\Sigma^{*+}$, $N^{*+}\left(\Delta^{*+}\right)$, $\bar{K}^{*0}$
intermediate states, the amplitudes are written as
\begin{equation}
    \mathcal{A}_{\lambda_{\gamma^{*}}, \lambda_{\lcm}, \lambda_p}^{\Sigma^{*+}}=\sum_{\lambda_{\Lambda_c^{+}}, \lambda_{\Sigma^{*+}}}A_{\lambda_{\gamma^{*}} ,\lambda_{\lcp}, \lambda_{\lcm}}^{\gamma^{*}\to\lcp\lcm} A_{\lambda_{\lcp},\lambda_{\Sigma^{*+}}}^{\lcp\to\Sigma^{*+}\piz} R_{\Sigma^{*+}}\left(M_{p\Ks}\right) A_{\lambda_{\Sigma^{*+}}, \lambda_{p}}^{\Sigma^{*+}\to p\Ks},
\end{equation}
\begin{align}
    \mathcal{A}_{\lambda_{\gamma^{*}}, \lambda_{\lcm}, \lambda_p}^{N^{*+}\left(\Delta^{*+}\right)}
     & = \sum_{\lambda_{\Lambda_c^{+}}, \lambda_{N^{*+}\left(\Delta^{*+}\right)}}
    A_{\lambda_{\gamma^{*}},\lambda_{\lcp}, \lambda_{\lcm}}^{\gamma^{*}\to\lcp\lcm}
    A_{\lambda_{\lcp},\lambda_{N^{*+}\left(\Delta^{*+}\right)}}^{\lcp\to N^{*+}\left(\Delta^{*+}\right)\Ks}
    \nonumber                                                                     \\
     & \quad \times R_{N^{*+}\left(\Delta^{*+}\right)}\left(M_{p\piz}\right)
    A_{\lambda_{N^{*+}\left(\Delta^{*+}\right)}, \lambda_{p}}^{N^{*+}\left(\Delta^{*+}\right)\to p\piz},
\end{align}
\begin{equation}
    \mathcal{A}_{\lambda_{\gamma^{*}}, \lambda_{\lcm}, \lambda_p}^{\bar{K}^{*0}}=\sum_{\lambda_{\Lambda_c^{+}}, \lambda_{\bar{K}^{*0}}} A_{\lambda_{\gamma^{*}} ,\lambda_{\lcp}, \lambda_{\lcm}}^{\gamma^{*}\to\lcp\lcm} A_{\lambda_{\lcp},\lambda_{p}, \lambda_{\bar{K}^{*0}}}^{\lcp\to p\bar{K}^{*0}} R_{\bar{K}^{*0}}\left(M_{\Ks\piz}\right)A_{\lambda_{\bar{K}^{*0}}}^{\bar{K}^{*0}\to \Ks\piz},
\end{equation}
where the amplitude \(A_{\lambda_{\gamma^{*}}, \lambda_{\Lambda_c^+}, \lambda_{\Lambda_c^-}}^{\gamma^{*}\to\Lambda_c^+\Lambda_c^-}\) is fixed according to the measured transverse polarization parameters reported in Ref.~\cite{Hongjian2023}.
Here, the propagator $R$ for the intermediate states $\Sigma^{*+}$,
$N^{*+}\left(\Delta^{*+}\right)$ and $\bar{K}^{*0}$ is chosen in the form of
the relativistic BW formula:
\begin{equation}
    R(m)=\frac{1}{m_0^2-m^2-i m_0 \Gamma(m)},
\end{equation}
with the mass-dependent width defined as:
\begin{equation}
    \Gamma(m)=\Gamma_0\left(\frac{q}{q_0}\right)^{2 l+1} \frac{m_0}{m} B_l^{\prime 2}\left(q, q_0, d\right),
\end{equation}
where $m_0$ and $\Gamma_0$ are the mass and width of each resonance.

For the decay through the $K \pi ~\mathcal{S}$-wave, the propagator is
described using the LASS model, with parameters fixed with the results reported
in Ref.~\cite{Adachi2018}. The full amplitude is the coherent sum of all
possible resonances, given as
\begin{equation}
    \begin{aligned}
        \mathcal{A}_{\lambda_{\gamma^{*}}, \lambda_{\lcm}, \lambda_p}
         & =  \mathcal{A}_{\lambda_{\gamma^{*}}, \lambda_{\lcm}, \lambda_p}^{\Sigma^{*+}}                                                                                                                                                                                                                                                                            \\
         & + \sum_{\lambda_{\lcm}^{\prime}, \lambda_p^{\prime}} \mathcal{A}_{\lambda_{\gamma^{*}}, \lambda_{\lcm}, \lambda_p^{\prime}}^{N^{*+}\left(\Delta^{*+}\right)} D_{\lambda_{\lcm}^{\prime}, \lambda_{\lcm}}^{1 / 2}\left(\alpha_{\lcm}, \beta_{\lcm}, \gamma_{\lcm}\right) D_{\lambda_p^{\prime}, \lambda_p}^{1 / 2}\left(\alpha_p, \beta_p, \gamma_p\right) \\
         & + \sum_{\lambda_{\lcm}^{\prime}, \lambda_p^{\prime}} \mathcal{A}_{\lambda_{\gamma^{*}}, \lambda_{\lcm}, \lambda_p^{\prime}}^{\bar{K}^{*0}} D_{\lambda_{\lcm}^{\prime}, \lambda_{\lcm}}^{1 / 2}\left(\alpha_{\lcm}, \beta_{\lcm}, \gamma_{\lcm}\right) D_{\lambda_p^{\prime}, \lambda_p}^{1 / 2}\left(\alpha_p, \beta_p, \gamma_p\right),                  \\
    \end{aligned}
\end{equation}
where the extra alignment $D$-functions are added to align final $\lcm$ and final proton states, correcting the helicities of $\lcm$ and proton in different chains. The details of the alignment angle calculations can be found in Ref.~\cite{Xinhai2022}.

The signal probability density function (PDF) for a given event is constructed using the full amplitude as
\begin{equation}
    P\left(x_i\right)=\frac{\varepsilon\left(x_i\right)\left|\mathcal{A}\left(x_i\right)\right|^2 R_4\left(x_i\right)}{\int \varepsilon\left(x_i\right)\left|\mathcal{A}\left(x_i\right)\right|^2 R_4\left(x_i\right) \mathrm{d} x_i},\quad|\mathcal{A}|^2= \sum_{\lambda_{\gamma^{*}}, \lambda_{\lcm}, \lambda_p}\left|\mathcal{A}_{\lambda_{\gamma^{*}}, \lambda_{\lcm}, \lambda_p}\right|^2,
    \label{eq:PDF}
\end{equation}
where $\varepsilon\left(x_i\right)$ is the detection efficiency parameterized in terms of the four-momenta of all final-state particles $x_i = (p_{1i}, p_{2i}, p_{3i}, p_{4i})$ for the $i$-th event, with $p_{1i}$, $p_{2i}$, $p_{3i}$, and $p_{4i}$ representing the four-momenta of $p$, $K_S^0$, $\piz$, and $\lcm$, respectively. $R_4\left(x_i\right)$ is the standard element of the four-body PHSP for the final state. The integration in the denominator is calculated using the PHSP MC sample mentioned in Sec.~\ref{sec:detector}:
\begin{equation}
    \int \varepsilon\left(x_i\right)\left|\mathcal{A}\left(x_i\right)\right|^2 R_4\left(x_i\right) \mathrm{d} x_i \propto \frac{1}{N_{\mathrm{PHSP}}} \sum_{i \in \mathrm{PHSP}}\left|\mathcal{A}\left(x_i\right)\right|^2.
    \label{eq:integration}
\end{equation}
The negative log-likelihood (NLL) is constructed using the background subtraction method:
\begin{equation}
    -\ln L=-\alpha\left[\sum_{i \in \mathrm{data}} \ln P\left(x_i\right)-w_{\mathrm{bkg}}^{\prime} \sum_{j \in \text { sideband }} \ln P\left(x_j\right)\right],
    \label{eq:NLL}
\end{equation}
where the background extrapolation factor $w_{\mathrm{bkg}}^{\prime}=w_{\mathrm{bkg}} \cdot N_{\text {data }} / N_{\text {sideband }}$, $w_{\mathrm{bkg}}$ is the background fraction, $ N_{\text {data }}$ and $N_{\text {sideband }}$ are the numbers of events in the $\mbc$ signal and sideband regions, respectively. To achieve an unbiased uncertainty estimate, a normalization factor~\cite{Kish1965} is taken into account, expressed as
\begin{equation}
    \alpha=\frac{N_{\text {data }}-N_{\text {sideband }} w_{\text {bkg }}^{\prime}}{N_{\text {data }}+N_{\text {sideband }} w_{\text {bkg }}^{\prime 2}}.
\end{equation}
The NLLs are constructed at each c.m. energy with shared parameters from the decays of $\lcp$ and resonance states, as well as the partial wave amplitudes $g_{ls}$ for the intermediate states in the process $\ee\to\lcp\lcm$. The joint NLL is obtained by summing all NLLs from each c.m. energy. The solution is found at the point where the joint NLL reaches its minimum value. 
The fit fractions (FFs) for each resonant component can be calculated as
\begin{equation}
    \mathrm{FF}_i=\frac{\int \left|\mathcal{A}_i\left(x_i\right)\right|^2 R_4\left(x_i\right) \mathrm{d} x_i}{\int \left|\sum_n \mathcal{A}_n\left(x_i\right)\right|^2 R_4\left(x_i\right) \mathrm{d} x_i},
\end{equation}
where $\mathcal{A}_i$ is the amplitude of the $i$-th component and the integration is calculated by the sum of truth level PHSP MC samples before requiring detector acceptance. Hence, the FFs of the interference part can be calculated as
\begin{equation}
    \mathrm{FF}_{i, j}=\frac{\int \left|\mathcal{A}_i\left(x_i\right)+\mathcal{A}_j\left(x_i\right)\right|^2 R_4\left(x_i\right) \mathrm{d} x_i}{\int \left|\sum_n \mathcal{A}_n\left(x_i\right)\right|^2 R_4\left(x_i\right) \mathrm{d} x_i}-\mathrm{FF}_i-\mathrm{FF}_j.
\end{equation}

\subsection{Nominal fit results}
\hspace{1.5em}

The starting point for constructing the nominal fit hypothesis in the amplitude
analysis is to include all possible excited $\Sigma^{*+}, \Delta^{*+}, N^{*+}$
and $\bar{K}^{*0}$ resonances that have experimental evidence of the existence
in the PDG \cite{ParticleDataGroup:2022pth}. The $\Delta(1600)^+$ is chosen as the reference channel, while the magnitude
and phase of its total amplitude are fixed to one and zero, respectively.

The statistical significance of these resonances is assessed by examining the
changes in the NLL and the number of degrees of freedom (nDOF) between the
models that include and exclude the individual component. In each iteration,
the resonance with the lowest significance is excluded if its significance is
below $5\sigma$. The process is then iterated, re-evaluating the significance
values of the remaining components, until only resonances with significances
exceeding $5\sigma$ are included. The final components identified in the
analysis include $\Delta(1600)^+$, $N(1710)^+$, $\bar{K}^{*}(892)^0$, and the
$K\pi~\mathcal{S}$-wave, with their statistical significances and parameter
configurations listed in Tab.~\ref{tab:model_components}. While the masses
and widths of $\Delta(1600)^+$ and $N(1710)^+$ are left free in the fit, the
corresponding parameters for $\bar{K}^{*}(892)^0$ are fixed to their PDG
values~\cite{ParticleDataGroup:2022pth}. The $K\pi~\mathcal{S}$-wave parameters
are fixed using the LASS model, with the values obtained from the
measurements in Ref. \cite{Adachi2018}.

\begin{table}[t]
    \centering
    \begin{tabular}{c|cccc} \hline\hline
        Component                 & Mass ($\mevcc$) & Width (MeV) & Model                  & Significance ($\sigma$) \\\hline
        $\Delta(1600)^+$          & Free         & Free     & Relativistic BW        & 14.5                    \\
        $N(1710)^+$               & Free         & Free     & Relativistic BW        & 6.2                     \\
        $\bar{K}^{*}(892)^0$      & 895.55          & 47.3        & Relativistic BW        & 12.8                    \\
        $K \pi ~\mathcal{S}$-wave & -               & -           & LASS~\cite{Adachi2018} & 11.3                    \\
        \hline\hline
    \end{tabular}
    \caption{Summary of the resonance parameter configurations, the models and the statistical significance. The parameters of $\Delta(1600)^+$ and $N(1710)^+$ are left free in the fit, while the parameters of $\bar{K}^{*}(892)^0$ are fixed to the PDG values. The $K\pi~\mathcal{S}$-wave parameters are fixed according to Ref.~\cite{Adachi2018}}.
    \label{tab:model_components}
\end{table}
\FloatBarrier

\begin{figure}[!t]
    \centering
    \includegraphics[width=0.32\textwidth]{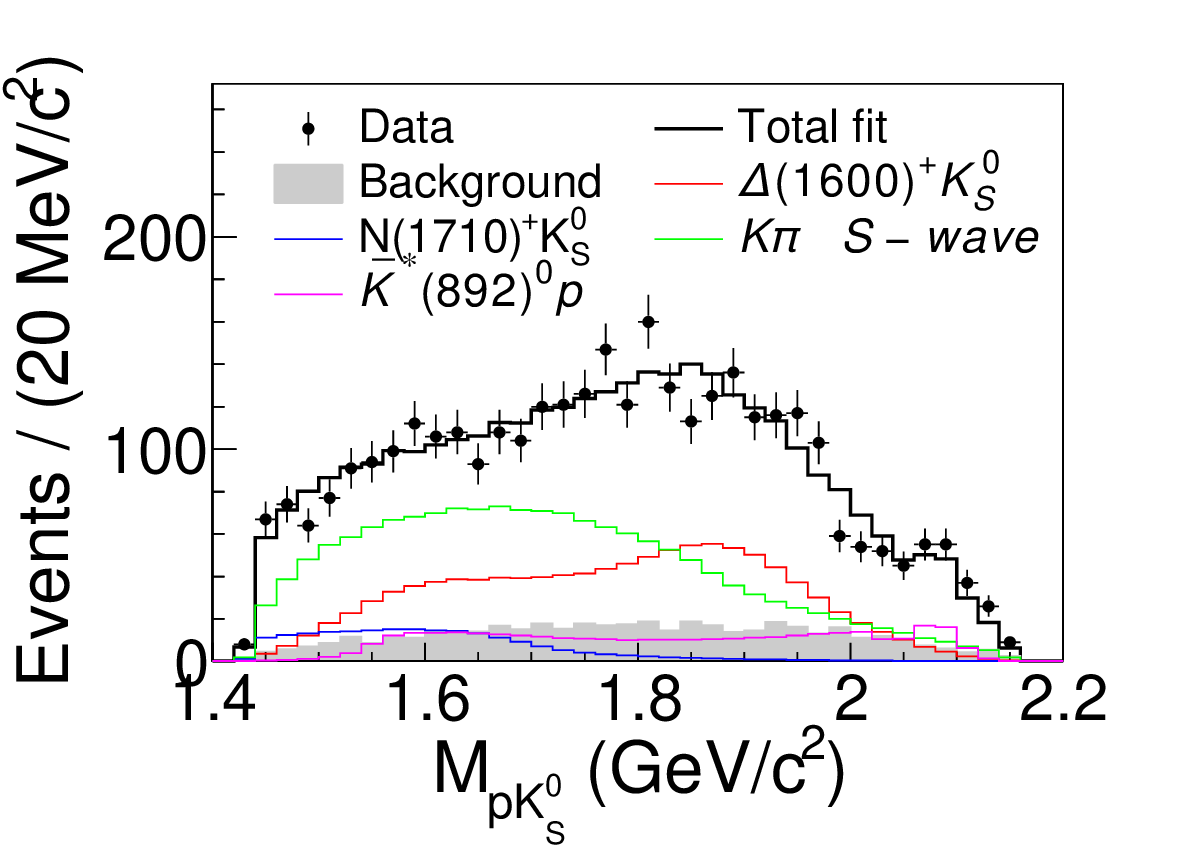}
    \includegraphics[width=0.32\textwidth]{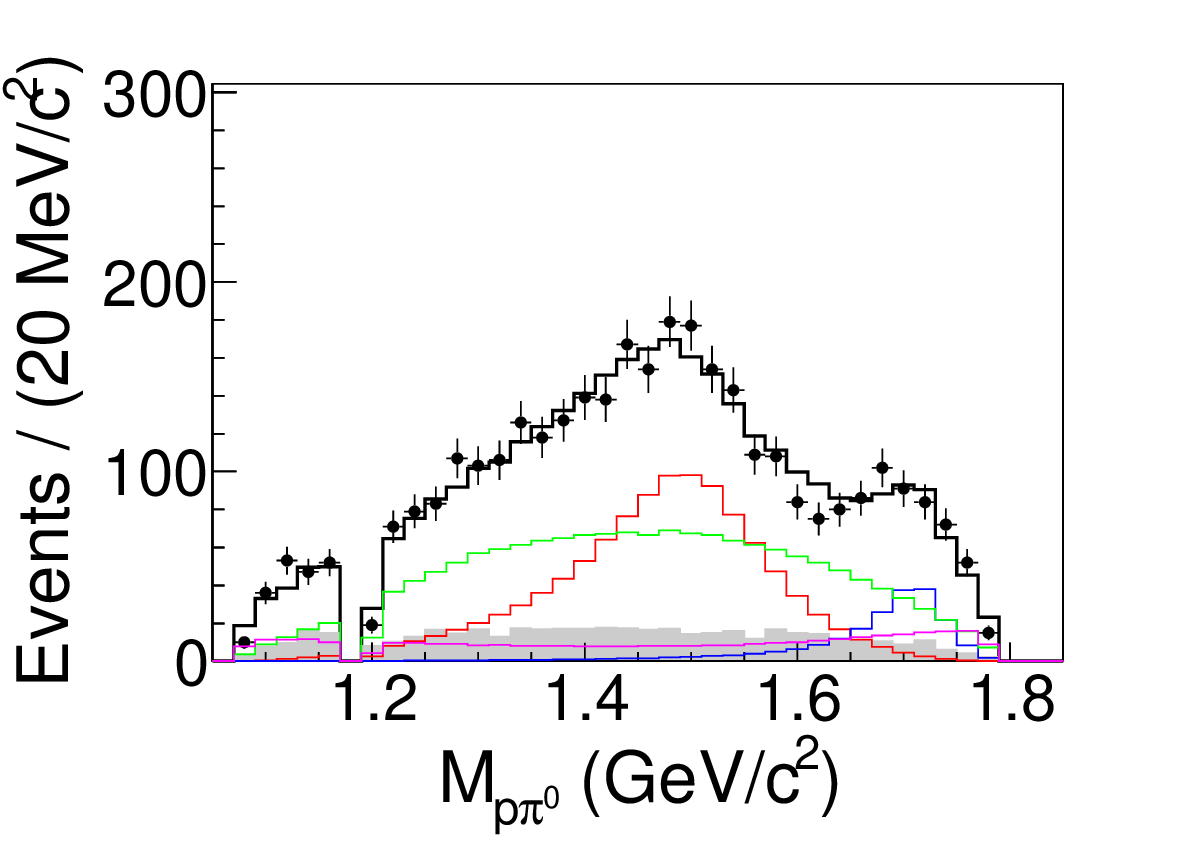}
    \includegraphics[width=0.32\textwidth]{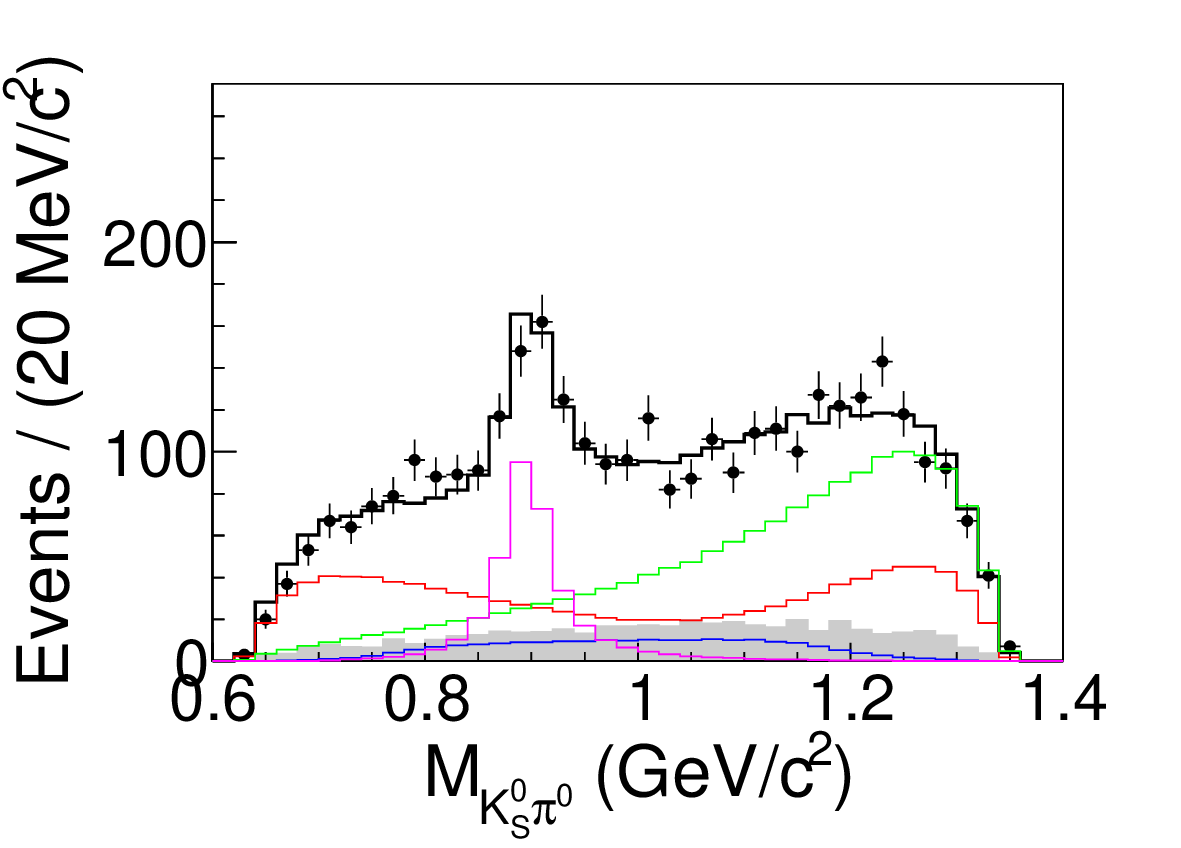}
    \caption{Projections of the fit results on (left) $M_{p\Ks}$, (middle) $M_{p\piz}$ and (right) $M_{\Ks\piz}$. Points with error bars denote data. The colored curves represent the different fit components, as indicated in the legend. }
    \label{fig:fit_mass_spectra}
\end{figure}

The projections of the fit onto the different invariant-mass spectra are shown in
Fig.~\ref{fig:fit_mass_spectra}. The mass and width for $\Delta(1600)^+$ are
$1520.4\pm 8.8~\mevcc$ and $199.4 \pm 25.6~\mev$.
For $N(1710)^+$, the mass and width are measured to be $1719.7 \pm 8.3~\mevcc$
and $72.0 \pm 19.6~\mev$, respectively. The results of FFs and interference
fractions (I.F.) are listed in Tab.~\ref{tab:Interaction_fractions}.

\begin{table}[!hpt]
    \centering
    \begin{tabular}{c|cccc}
        \hline\hline
        Component                 & $\Delta(1600)^{+}$ & $\bar{K}^{*}(892)^0$ & $K \pi ~\mathcal{S}$-wave & $N(1710)^{+}$ \\
        \hline
        $\Delta(1600)^{+}$        & $37.0 \pm 3.5$     & -                    & -                         & -             \\
        $\bar{K}^{*}(892)^0$      & $-2.2 \pm 1.5$     & $12.0 \pm 2.0$       & -                         & -             \\
        $K \pi ~\mathcal{S}$-wave & $-6.7 \pm 1.9$     & $-0.0 \pm 0.0$       & $53.8 \pm 4.7$            & -             \\
        $N(1710)^{+}$             & $-0.0 \pm 0.0$     & $2.2 \pm 0.9$        & $-3.3 \pm 2.4$            & $7.2 \pm 2.2$ \\
        \hline\hline
    \end{tabular}
    \caption{Summary of FFs on the diagonal and I.F. on the off-diagonal, both expressed in units of $10^{-2}$. The uncertainties are statistical only.}
    \label{tab:Interaction_fractions}
\end{table}

\section{Systematic uncertainties}
\label{sec:systematic}
\hspace{1.5em} The systematic uncertainties of the measured observables, including the masses and widths of the $\Delta(1600)^{+}$ and $N(1710)^{+}$ resonances, as well as the FFs of the $\Delta(1600)^{+}$, $N(1710)^{+}$, and $\bar{K}^{*}(892)^0$ components, are estimated.  In general, the following sources are considered: (I) fixed parameters, (II) resonance components, (III) radius parameter, (IV) polarization of the initial $\lcp$, (V) fit bias, (VI) data-MC differences, and (VII) background description. All systematic uncertainties are listed in Tab.~\ref{tab:sysU_summary}.

\begin{table}[t]
    \centering
    \small
    \begin{tabular}{c|c|ccccccc|c}
        \hline\hline
        \multirow{2}{*}{Parameter}                     & \multirow{2}{*}{Stat.} & \multicolumn{7}{c|}{Syst. Source} & \multirow{2}{*}{Total Syst.}                                       \\
        \cline{3-9}
                                                       &                        & I                                 & II                           & III  & IV  & V   & VI  & VII &      \\
        \hline
        $\mathrm{Mass~of~}\Delta(1600)^{+}$ ($\mevcc$) & 8.8                    & 0.4                               & 34.8                         & 13.8 & 0.3 & 2.6 & 2.1 & 3.0 & 37.7 \\
        \rule{0pt}{15pt}
        $\mathrm{Width~of~}\Delta(1600)^{+}$ ($\mev$)  & 25.6                   & 3.0                               & 37.4                         & 9.4  & 0.9 & 2.1 & 9.4 & 2.2 & 39.9 \\
        \rule{0pt}{15pt}
        $\mathrm{Mass~of~}N(1710)^{+}$ ($\mevcc$)      & 8.9                    & 0.2                               & 8.8                          & 1.9  & 0.4 & 0.8 & 0.4 & 0.3 & 9.1  \\
        \rule{0pt}{15pt}
        $\mathrm{Width~of~}N(1710)^{+}$ ($\mev$)       & 19.6                   & 1.0                               & 13.2                         & 1.5  & 0.8 & 2.1 & 6.4 & 0.4 & 15.0 \\
        \rule{0pt}{15pt}
        $\mathrm{FFs~of~}\Delta(1600)^{+}$ (\%)        & 3.5                    & 0.3                               & 25.3                         & 1.2  & 0.2 & 0.1 & 0.3 & 0.9 & 25.4 \\
        \rule{0pt}{15pt}
        $\mathrm{FFs~of~}N(1710)^{+}$   (\%)           & 2.2                    & 0.2                               & 2.7                          & 0.6  & 0.2 & 0.7 & 1.7 & 1.1 & 3.5  \\
        \rule{0pt}{15pt}
        $\mathrm{FFs~of~}\bar{K}^{*}(892)^0$ (\%)      & 2.0                    & 0.1                               & 0.9                          & 0.5  & 0.2 & 0.3 & 0.4 & 0.6 & 1.3  \\
        \hline\hline
    \end{tabular}
    \caption{Absolute systematic uncertainties (values are in the same units as the corresponding statistical uncertainties) on the masses and widths of $\Delta(1600)^{+}$ and $N(1710)^{+}$, as well as the FFs of the $\Delta(1600)^{+}$, $N(1710)^{+}$, and $\bar{K}^{*}(892)^0$. The columns Stat. and Syst. denote statistical and systematic uncertainties, respectively. The total systematic uncertainties are calculated by summing up all contributions in quadrature.}
    \label{tab:sysU_summary}
\end{table}

\begin{enumerate}[I]

    \item \textbf{Fixed parameters}. In the nominal fit, the mass and width for $\bar{K}^{*}(892)^0$ are fixed. The $K \pi ~\mathcal{S}$-wave is fixed using the parameters of the LASS model~\cite{Adachi2018}. To estimate the corresponding systematic uncertainties, each parameter is individually varied by $\pm 1 \sigma$ one at a time, while keeping the other parameters fixed, and the fit procedure is repeated. The largest absolute deviations from the nominal fit results are assigned as individual systematic uncertainties.

    \item \textbf{Resonance components}. To evaluate the systematic uncertainty associates to the choice of the resonant components in the amplitude model, we investigate the impact of including the $N(1535)^+$ state, which has the highest statistical significance ($4.1\sigma$) among the excluded resonances and exhibits strong kinematic overlap with the $\Delta(1600)^+$ in the $p\pi^0$ spectrum. The nominal fit excludes $N(1535)^+$ as its significance falls below our $5\sigma$ inclusion threshold. However, recognizing that this criterion, while objective, may not fully capture potential model ambiguities, we perform an alternative five-component fit that includes $N(1535)^+$ in addition to the nominal components. The resulting variations in the fitted parameters, particularly for the $\Delta(1600)^+$ whose mass, width, and fit fraction are most sensitive to this addition, are taken as the systematic uncertainty associated with resonance modeling. The impact on the $N(1710)^+$ and $\bar{K}^{*}(892)^0$ parameters is found to be significantly smaller. The detailed results of the five-component fit and its implications are discussed in Sec.~\ref{sec:discussion_model_dep}, where they are used to assess the model dependence.

    \item \textbf{Radius parameter}. In the nominal fit, the radius parameter $d$ is chosen as $d=$ $0.73~\text{fm}$ \cite{Xinhai2022}. To estimate the associated systematic uncertainty, the fits are performed by setting $d$ at the alternative values derived from Refs. \cite{Xinhai2022, BESIII:2020rad}, $d=0.53~\text{fm}$ and $d=1.16~\text{fm}$, and the largest variations are assigned as the systematic uncertainties.

    \item \textbf{Polarization of the initial $\lcp$}. To evaluate the systematic uncertainty associated with the fixed $\lcp$ transverse polarization parameters, several independent samples of polarization parameters are generated from a multivariate normal distribution, incorporating the correlations of Ref. \cite{Hongjian2023}. Amplitude analysis fits are performed for each sample, propagating the parameter uncertainties to the residuals between the fit and the nominal results. The Gaussian widths of the residual distributions are assigned as the corresponding systematic uncertainties.

    \item \textbf{Fit bias}. To evaluate the potential fit bias, an input-output check is performed using toy MC samples generated from the nominal amplitude model with the best-fit parameters, each with the same statistics as the data. The same fit procedure is applied to each toy sample, and the pull value is calculated as the difference between the toy and the nominal results, defined as:
          \begin{equation}
              \text{pull} = \frac{\text{output value} - \text{input value}}{\text{output uncertainty}}.
          \end{equation}
          The pull distributions are fitted with a Gaussian function. The resulting mean $(\mu_{\text{pull}})$ and standard deviation $(\sigma_{\text{pull}})$ are then used to compute bias-corrected reference results:
          \begin{equation}
              \begin{aligned}
                  \sigma_{\text{ref}} & = \sigma_{\text{nom}} \times \sigma_{\text{pull}},                 \\
                  \mu_{\text{ref}}    & = \mu_{\text{nom}} - \mu_{\text{pull}} \times \sigma_{\text{ref}},
              \end{aligned}
              \label{eq:correction}
          \end{equation}
          where $\mu_{\text{nom}}$ and $\sigma_{\text{nom}}$ are the nominal fit results. The fitted mean $(\mu_{\text{pull}})$, variation $(\sigma_{\text{pull}})$ values, and reference results are summarised in Tab.~\ref{tab:Fit_bias}. The systematic uncertainties are calculated as the differences between the nominal and the reference results.

          \begin{table}[t]
              \centering

              \begin{tabular}{c|cc|c}
                  \hline
                  \hline Parameter                       & $\mu_{\text {pull }}$ & $\sigma_{\text {pull }}$ & Reference result                   \\
                  \hline
                  $\mathrm{Mass ~of~} \Delta(1600)^{+}$  & $-0.21 \pm 0.05$      & $1.00 \pm 0.04$          & $1522.2\pm 8.8 ~\mathrm{(\mevcc)}$ \\
                  $\mathrm{Width ~of~} \Delta(1600)^{+}$ & $0.17 \pm 0.06$       & $1.03 \pm 0.05$          & $194.9 \pm 26.4 ~\mathrm{(\mev)}$  \\
                  $\mathrm{Mass ~of~} N(1710)^{+}$       & $-0.22 \pm 0.06$      & $1.13 \pm 0.05$          & $1721.8\pm 9.4 ~\mathrm{(\mevcc)}$ \\
                  $\mathrm{Width ~of~} N(1710)^{+}$      & $-0.23 \pm 0.09$      & $0.92 \pm 0.06$          & $76.1 \pm 18.0 ~\mathrm{(\mev)}$   \\
                  \hline
                  $\mathrm{FFs ~of~} \Delta(1600)^{+}$   & $0.24 \pm 0.05$       & $1.04 \pm 0.05$          & $(36.1 \pm 3.6)\%$                 \\
                  $\mathrm{FFs ~of~} N(1710)^{+}$        & $-0.14 \pm 0.05$      & $0.98 \pm 0.04$          & $(7.5 \pm 2.2)\%$                  \\
                  $\mathrm{FFs ~of~} \bar{K}^{*}(892)^0$ & $0.20 \pm 0.06$       & $1.07 \pm 0.04$          & $(11.6 \pm 2.1)\%$                 \\
                  \hline\hline
              \end{tabular}
              \caption{Fit Bias uncertainty:  fitted mean $(\mu_{pull})$, variance $(\sigma_{pull})$ and reference results. The uncertainties are statistical only.}
              \label{tab:Fit_bias}
          \end{table}

    \item \textbf{Efficiency estimation}. To estimate possible systematic uncertainties due to the detection efficiency differences between data and MC simulations, the effects from the tracking of the proton and $\pip$ candidates, PID of the proton candidates, the $\pi^0$ reconstruction and $\Ks$ reconstruction, as well as the data-MC differences from the input variables of the BDTG are considered.

          The differences in reconstruction efficiency between data and MC simulations have been investigated in earlier BESIII studies, specifically for $\piz$ reconstruction in Ref.~\cite{BESIII:2017pi0}, and for tracking and PID efficiencies of proton and $\pip$ in Ref.~\cite{BESIII:2022bkj}. For $\Ks$ reconstruction, the $J/\psi\to\Ks
              K^{\pm} \pi^{\mp}$ decay is chosen as control sample to study the $\Ks$
          reconstruction efficiency using our loose selection criteria documented in
          Sec.~\ref{sec:eventSelection}. The correction factors
          $w=\varepsilon_{\mathrm{Data}} / \varepsilon_{\mathrm{MC}}$ are assigned as the
          weighting factors of the PHSP MC sample and the fit procedure is repeated.
          Including the weighting factors, eq. \ref{eq:integration} becomes
          \begin{equation}
              \int \varepsilon\left(p_j\right)\left|\mathcal{A}\left(p_j\right)\right|^2 R_4\left(p_j\right) \mathrm{d} p_j \propto \frac{1}{N_{\mathrm{PHSP}}} \sum_{i \in \mathrm{PHSP}}w\left(x_j\right)\left|\mathcal{A}\left(x_i\right)\right|^2,
              \label{eq:integration2}
          \end{equation}
          where $w(x_i)$ is the weighting factor for the $i$-th event. The resulting variations in the final results are considered as the systematic uncertainties.

          The discrepancies between data and MC simulations introduced by the
          BDTG score cut arise from inconsistencies in the input variables for the model. We
          assume that after correcting the input variable distributions, the application
          of the nominal BDTG cut does not introduce additional discrepancies. To
          estimate the associated systematic uncertainties, unified weights for the
          signal MC sample are determined. This weight are derived iteratively to
          achieve optimal consistency between the signal MC and data distributions across
          all eight input variables simultaneously. The resulting signal weights are then
          applied to the signal MC sample. This re-weighted sample is used during the
          construction of the new amplitudes, as described in eq. \ref{eq:integration2}.
          With the BDTG cut maintained consistently, the fit procedure is repeated. The
          variations in the results are considered as the systematic uncertainties.

    \item \textbf{Background description}. The systematic uncertainties associated with the background description can be classified into two categories: background fractions and background shape.

          In the nominal fit, the background fractions are fixed according to the fit
          results given in Tab.~\ref{tab:BDTG_MBc_Scan}. To estimate the systematic
          uncertainties, the background fraction for each c.m. energy is varied within
          $\pm 1\sigma$, and the fit procedure is repeated. The quadratic sum of the
          largest variations in each fraction are assigned as the associated systematic
          uncertainties.

          The background is modeled using the \(M_{\text{BC}}\) sideband region of the
          data to describe the shape in the signal region. To study the possible bias
          introduced by this approach, an investigation is performed using the inclusive
          MC sample to examine the consistency between the \(M_{\text{BC}}\) sideband and
          signal regions. Based on the observed differences in the MC modeling, the
          shape of the \(M_{\text{BC}}\) sideband in data is adjusted accordingly, and the
          fit procedure is repeated. The resulting variations in the final results are
          taken as the systematic uncertainties.
\end{enumerate}

The total systematic uncertainty is obtained by summing up all contributions in
quadrature, as listed in Tab.~\ref{tab:sysU_summary}.

\section{Discussion on the $N(1535)^+$ contribution}
\label{sec:discussion_model_dep}
\hspace{1.5em} The nominal model is constructed based on a predefined \(5\sigma\) statistical significance threshold, which excludes the \(N(1535)^+\) resonance. However, this resonance has a sub-threshold significance of $4.1\sigma$ (see Tab.~\ref{tab:significance_alternative}) and significantly influences the $\Delta(1600)^+$ parameters. Therefore, the stability of the model requires a focused discussion.

\begin{table}[t]
    \centering
    \begin{tabular}{c|ccccc}
        \hline\hline
        Component               & $\Delta(1600)^+$ & $N(1710)^+$ & $N(1535)^+$ & $\bar{K}^{*}(892)^0$ & $K \pi ~\mathcal{S}$-wave \\
        \hline
        Significance ($\sigma$) & 7.0              & 5.7         & 4.1         & 11.8                 & 10.9                      \\
        \hline\hline
    \end{tabular}
    \caption{Statistical significances of components in the alternative fit including $N(1535)^+$.}
    \label{tab:significance_alternative}
\end{table}

\begin{figure}[!t]
    \centering
    \includegraphics[width=0.32\textwidth]{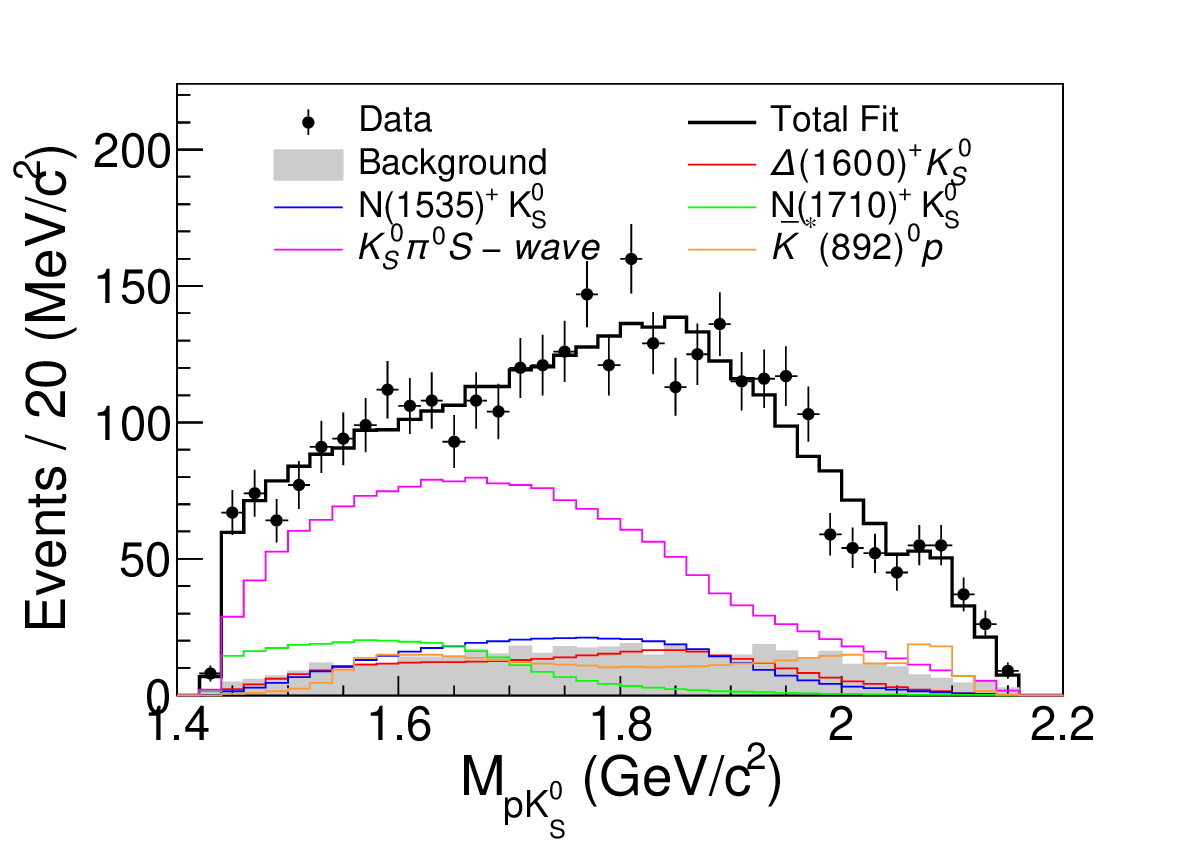}
    \includegraphics[width=0.32\textwidth]{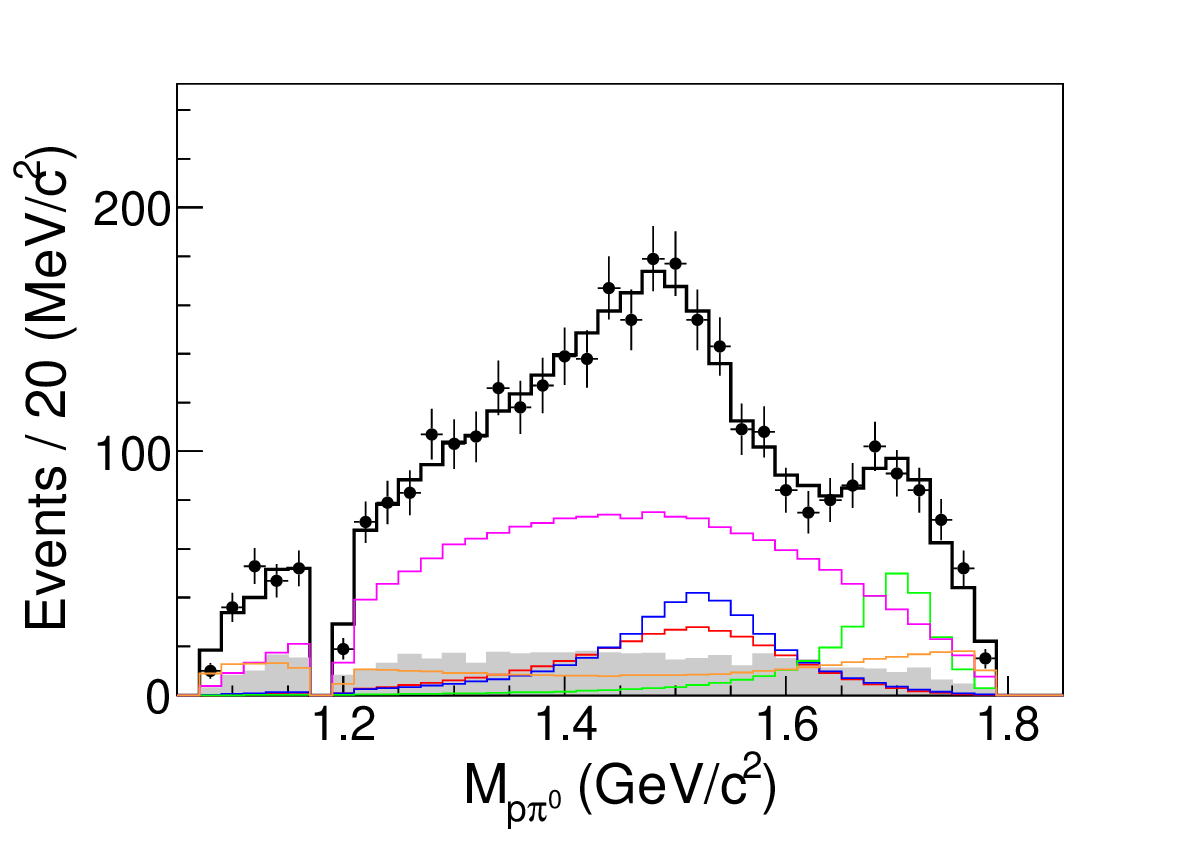}
    \includegraphics[width=0.32\textwidth]{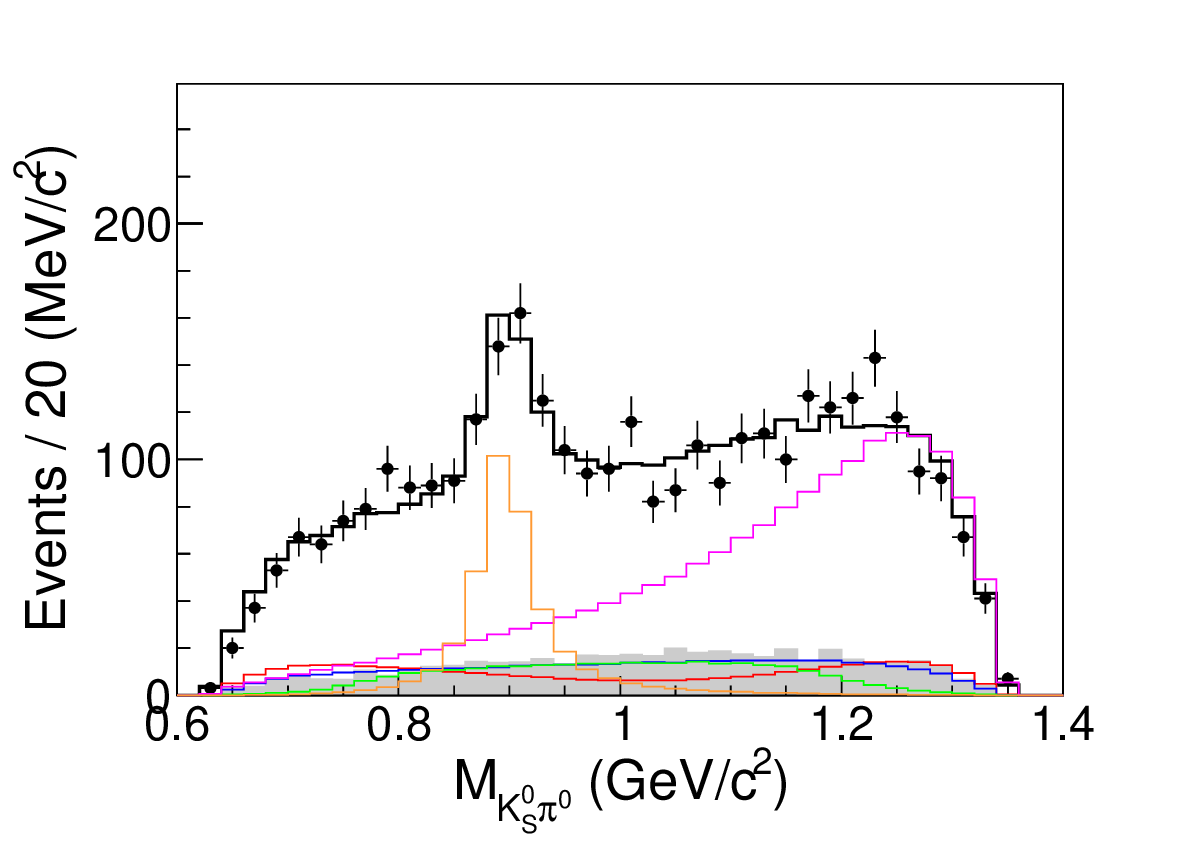}
    \caption{Projections of the alternative fit results on (left) $M_{p\Ks}$, (middle) $M_{p\piz}$ and (right) $M_{\Ks\piz}$. Points with error bars denote data. The colored curves represent the different fit components, as indicated in the legend.}
    \label{fig:fit_mass_spectra_alternative}
\end{figure}

\begin{table}[t]
    \centering
    \begin{tabular}{l|cc}
        \hline\hline
        Parameter                           & Nominal Model    & Model with $N(1535)^+$ \\
        \hline
        Mass of $\Delta(1600)^+$ ($\mevcc$) & $1520.4 \pm 8.8$ & $1555.2 \pm 34.3$      \\
        \rule{0pt}{15pt}
        Width of $\Delta(1600)^+$ ($\mev$)  & $199.4 \pm 26.4$ & $236.8 \pm 77.5$       \\
        \rule{0pt}{15pt}
        FF of $\Delta(1600)^+$ (\%)         & $37.0 \pm 3.5$   & $11.7 \pm 4.1$         \\
        \hline
        \rule{0pt}{15pt}
        Mass of $N(1710)^+$ ($\mevcc$)      & $1719.7 \pm 9.4$ & $1710.9 \pm 12.2$      \\
        \rule{0pt}{15pt}
        Width of $N(1710)^+$ ($\mev$)       & $72.0 \pm 18.0$  & $85.3 \pm 23.1$        \\
        \rule{0pt}{15pt}
        FF of $N(1710)^+$ (\%)              & $7.2 \pm 2.2$    & $9.9 \pm 2.8$          \\
        \rule{0pt}{15pt}
        FF of $\bar{K}^{*}(892)^0$ (\%)     & $12.0 \pm 2.0$   & $12.9 \pm 3.0$         \\
        \hline
        \rule{0pt}{15pt}
        FF of $N(1535)^+$ (\%)              & -                & $13.7 \pm 7.4$         \\
        \hline\hline
    \end{tabular}
    \caption{Comparison of the key parameters between the nominal fit and the alternative model including $N(1535)^+$. Only statistical uncertainties are shown.}
    \label{tab:comparison_model_dep}
\end{table}

Figure~\ref{fig:fit_mass_spectra_alternative} demonstrates that the five-component model, which includes \(N(1535)^+\), also provides a reasonable description of the data. The key outcome is summarized in Tab.~\ref{tab:comparison_model_dep}. While the parameters of the \(N(1710)^+\) and \(\bar{K}^{*}(892)^0\) remain stable within their respective uncertainties, those of the \(\Delta(1600)^+\) change significantly. Its fit fraction decreases from \((37.0 \pm 3.5)\%\) to \((11.7 \pm 4.1)\%\), and its extracted width becomes \(\Gamma = 236.8 \pm 77.5~\mev\), exhibiting a significantly larger statistical uncertainty compared to the nominal result.

The observed behavior originates from the limited statistical precision of the current data sample.
Both the \(\Delta(1600)^+\) and \(N(1535)^+\) are broad resonances with comparable masses and large widths, and both decay to the \(p\pi^0\) final state. This leads to a significant kinematic overlap in the \(p\pi^0\) invariant mass spectrum, as is visible in the middle panels of Fig.~\ref{fig:fit_mass_spectra} and~\ref{fig:fit_mass_spectra_alternative}. With the present statistics, the experimental measurement is not yet sufficient to disentangle the \(p\pi^0\) spectrum and uniquely assign its contributions to these two specific resonances. Consequently, when both components are introduced into the model simultaneously, a strong anti-correlation between their amplitudes arises. This manifests itself as large variations in the fit results and substantially increased uncertainties, as an increase in the contribution of one component is compensated by a decrease in the other.

Therefore, we adopt the nominal four-component model as our primary result, as it is defined by an objective selection criterion and yields well-constrained parameters. To account for the ambiguity in distinguishing the contributions of \(\Delta(1600)^+\) and \(N(1535)^+\) with the present data sample, the parameter variations resulting from the alternative model that includes \(N(1535)^+\) are evaluated and included as a major source of systematic uncertainty (Source II in Tab.~\ref{tab:sysU_summary}). A definitive resolution of this ambiguity will require future experimental data with higher statistics, potentially through a combined analysis of multiple decay channels.

\section{Summary}
\label{sec:summary}
\hspace{1.5em} In summary, based on $\ee$ annihilation data at c.m. energies between 4.600 and 4.699 GeV corresponding to an integrated luminosity of 4.5 $\text{fb}^{-1}$, the first amplitude analysis of $\lcp\to p\Ks\piz$ is performed. From the analysis, the masses and widths of $\Delta(1600)^+$ and $N(1710)^{+}$ are determined to be:
\begin{equation}
    m_{\Delta(1600)^{+}} = (1520.4\pm8.8\pm37.7) ~\mevcc,
\end{equation}
\begin{equation}
    \Gamma_{\Delta(1600)^{+}} = (199.4\pm26.4\pm39.9) ~\mev,
\end{equation}
\begin{equation}
    m_{N(1710)^{+}} = (1719.7 \pm 9.4  \pm 9.1) ~\mevcc,
\end{equation}
\begin{equation}
    \Gamma_{N(1710)^{+}} = (72.0 \pm 18.0 \pm 15.0) ~\mev.
\end{equation}
The measured masses and widths of the $\Delta(1600)^+$ and $N(1710)^{+}$ are consistent with the PDG values~\cite{ParticleDataGroup:2022pth}. The FFs for three two-body processes are measured. Using the recent BESIII measurement $\mathcal{B}\left(\Lambda_c^{+} \rightarrow p K_{\mathrm{S}}^0 \pi^0\right) = (2.19 \pm 0.06 \pm 0.05) \%$~\cite{Liuying2025}, the product BFs are obtained to be:
\begin{equation}
    \mathcal{B}\left(\Lambda_c^{+} \rightarrow \Delta(1600)^{+} K_S^0\right) \cdot \mathcal{B}\left(\Delta(1600)^{+} \rightarrow p \pi^0\right) = \left(8.10 \pm 0.80 \pm 5.57\right) \times 10^{-3},
\end{equation}
\begin{equation}
    \mathcal{B}\left(\Lambda_c^{+} \rightarrow N(1710)^{+} K_S^0\right) \cdot \mathcal{B}\left(N(1710)^{+} \rightarrow p \pi^0\right) = \left(1.58 \pm 0.48 \pm 0.77\right) \times 10^{-3},
\end{equation}
\begin{equation}
    \mathcal{B}\left(\Lambda_c^{+} \rightarrow p \bar{K}^{*}(892)^0\right) \cdot \mathcal{B}\left(\bar{K}^{*}(892)^0 \rightarrow K_S^0 \pi^0\right) = \left(2.63 \pm 0.47 \pm 0.29\right) \times 10^{-3},
\end{equation}
where the first uncertainties are statistical and the second systematic. The statistical uncertainties have been corrected to account for potential biases that are estimated from the input-output discrepancy, as listed in Tab.~\ref{tab:Fit_bias}.

Based on isospin symmetry and the assumption that $\bar{K}^{*}(892)^0$ decays entirely to $K\pi$ final states, we derive $\mathcal{B}(\Lambda_c^{+} \rightarrow p \bar{K}^{*}(892)^0) = (1.58 \pm 0.33) \times 10^{-2}$. 
For comparison, using the LHCb measurement $\mathcal{B}(\bar{K}^{*}(892)^0 \to K^-\pi^+) = (22.14 \pm 0.23 \pm 0.04 \pm 0.64)\%$~\cite{PhysRevD.108.012023} and $\mathcal{B}(\Lambda_c^+ \to p K^- \pi^+) = (6.61 \pm 0.11 \pm 0.12)\%$~\cite{Liuying2025} with the isospin factor $2/3$, we obtain $\mathcal{B}(\Lambda_c^{+} \rightarrow p \bar{K}^{*}(892)^0) = (2.20 \pm 0.09) \times 10^{-2}$. 
Our result is consistent with the LHCb-based value, with a deviation of $1.8\sigma$.

A key finding of this analysis is the significant model dependence observed between the contributions of the broad, overlapping resonances $\Delta(1600)^+$ and $N(1535)^+$ to the $p\pi^0$ spectrum. This ambiguity highlights an important source of uncertainty in extracting the parameters of such resonances from a single decay channel with the present statistics. Future high-precision measurements or combined multi-channel analyses will be essential to uniquely determine their individual contributions.

In conclusion, this study provides new measurements for the properties of $\Delta(1600)^+$, $N(1710)^+$, and $\bar{K}^{*}(892)^0$ through the amplitude analysis of $\Lambda_c^+ \to p K_S^0 \pi^0$. More importantly, it reveals the common challenge of disentangling broad, overlapping resonances in charmed baryon decays. These results offer important insights into the weak decay mechanisms of $\Lambda_c^+$, the nature of baryonic excited states, and hadronic final-state interactions, thereby contributing valuable experimental input to this field.

\acknowledgments
\hspace{1.5em}
The BESIII Collaboration thanks the staff of BEPCII (https://cstr.cn/31109.02.BEPC) and the IHEP computing center for their strong support. This work is supported in part by National Key R\&D Program of China under Contracts Nos. 2023YFA1606000, 2023YFA1606704, 2025YFA1613900; National Natural Science Foundation of China (NSFC) under Contracts Nos. 11635010, 11935015, 11935016, 11935018, 12025502, 12035009, 12035013, 12061131003, 12192260, 12192261, 12192262, 12192263, 12192264, 12192265, 12221005, 12225509, 12235017, 12342502, 12361141819, 12535005; the Chinese Academy of Sciences (CAS) Large-Scale Scientific Facility Program; the Strategic Priority Research Program of Chinese Academy of Sciences under Contract No. XDA0480600; CAS under Contract No. YSBR-101; 100 Talents Program of CAS; The Institute of Nuclear and Particle Physics (INPAC) and Shanghai Key Laboratory for Particle Physics and Cosmology; Agencia Nacional de Investigaci\'on y Desarrollo de Chile (ANID), Chile under Contract No. ANID CCTVal CIA250027; Istituto Nazionale di Fisica Nucleare, Italy; Knut and Alice Wallenberg Foundation under Contracts Nos. 2021.0174, 2021.0299, 2023.0315; Ministry of Development of Turkey under Contract No. DPT2006K-120470; National Research Foundation of Korea under Contract No. RS-2026-25486791; National Science and Technology fund of Mongolia; Polish National Science Centre under Contracts Nos. 2024/53/B/ST2/00975, 2025/58/E/ST2/00051; STFC (United Kingdom); Swedish Research Council under Contract No. 2019.04595; Thailand NSRF via the Research and Innovation Acceleration Agency for Competitiveness and Area Development (RCAD) (Program Management Unit for Frontier Brainpower and Future Industries) under Contract No. B39G690072; U. S. Department of Energy under Contract No. DE-FG02-05ER41374
\bibliographystyle{JHEP}
\bibliography{main.bib}
\newpage

\newpage
\section*{The BESIII collaboration}
\addcontentsline{toc}{section}{The BESIII collaboration}
\begin{small}
    \begin{center}
        M.~Ablikim$^{1}$\BESIIIorcid{0000-0002-3935-619X},
        M.~N.~Achasov$^{4,c}$\BESIIIorcid{0000-0002-9400-8622},
        P.~Adlarson$^{88}$\BESIIIorcid{0000-0001-6280-3851},
        X.~C.~Ai$^{95}$\BESIIIorcid{0000-0003-3856-2415},
        C.~S.~Akondi$^{33A,33B}$\BESIIIorcid{0000-0001-6303-5217},
        R.~Aliberti$^{42}$\BESIIIorcid{0000-0003-3500-4012},
        A.~Amoroso$^{87A,87C}$\BESIIIorcid{0000-0002-3095-8610},
        L.~P.~An$^{54,h}$\BESIIIorcid{0000-0002-3274-5627},
        Q.~An$^{83,70,\dagger}$,
        M.~S.~Anderson$^{42}$\BESIIIorcid{0009-0008-1550-2632}
% ...
        \\
        \vspace{0.2cm}
        (BESIII Collaboration)\\
        \vspace{0.2cm} {\it
            $^{1}$ Institute of High Energy Physics, Beijing 100049, People's Republic of China\\
% ...
$^{4}$ Budker Institute of Nuclear Physics SB RAS (BINP), Novosibirsk 630090, Russia\\
% ...
$^{33}$ INFN Sezione di Ferrara, (A)INFN Sezione di Ferrara, I-44122, Ferrara, Italy; (B)University of Ferrara, I-44122, Ferrara, Italy\\
% ...
$^{42}$ Johannes Gutenberg University of Mainz, Johann-Joachim-Becher-Weg 45, D-55099 Mainz, Germany\\
% ...
$^{54}$ Peking University, Beijing 100871, People's Republic of China\\
% ...
$^{70}$ State Key Laboratory of Particle Detection and Electronics, Beijing 100049, Hefei 230026, People's Republic of China\\
% ...
$^{83}$ University of Science and Technology of China, Hefei 230026, People's Republic of China\\
% ...
$^{87}$ University of Turin and INFN, (A)University of Turin, I-10125, Turin, Italy; (B)University of Eastern Piedmont, I-15121, Alessandria, Italy; (C)INFN, I-10125, Turin, Italy\\
            $^{88}$ Uppsala University, Box 516, SE-75120 Uppsala, Sweden\\
% ...
$^{95}$ Zhengzhou University, Zhengzhou 450001, People's Republic of China\\

            \vspace{0.2cm}
            $^{\dagger}$ Deceased\\
% ...
$^{c}$ Also at the Novosibirsk State University, Novosibirsk, 630090, Russia\\
% ...
$^{h}$ Also at State Key Laboratory of Nuclear Physics and Technology, Peking University, Beijing 100871, People's Republic of China\\
% ...
        }\end{center}

    \vspace{0.4cm}
\end{small}
\end{document}